\pdfoutput=1 
\documentclass[12pt]{iopart}
\usepackage{setstack}
\usepackage{graphicx,amssymb,amsthm,epsfig,amsbsy,amsfonts,mathrsfs}
\usepackage[english,french]{babel}

\usepackage[T1]{fontenc}

\usepackage{bbm}

\usepackage{url,hyperref}                            

\usepackage{cite}

\usepackage{xcolor}

\definecolor{Cblue}{HTML}{045FB4}
\definecolor{Cred}{HTML}{DF0101}
\definecolor{CDarkred}{HTML}{8A0808}

\hypersetup{
    colorlinks=true, 
    linktoc=all,     
    linkcolor=black,
    citecolor=Cblue
  }

\usepackage{dsfont}
\def\un{\mathds{1}}

\def\Xint#1{\mathchoice
  {\XXint\displaystyle\textstyle{#1}}%
  {\XXint\textstyle\scriptstyle{#1}}%
  {\XXint\scriptstyle\scriptscriptstyle{#1}}%
  {\XXint\scriptscriptstyle\scriptscriptstyle{#1}}%
  \!\int}
\def\XXint#1#2#3{{\setbox0=\hbox{$#1{#2#3}{\int}$}
    \vcenter{\hbox{$#2#3$}}\kern-.5\wd0}}

\def\intVp{\Xint-}

\newcommand{\dt}[2]{\ensuremath{\frac{\dd #1}{\dd #2}}}
\newcommand{\dtn}[3]{\ensuremath{\frac{\dd^{#3} #1}{\dd #2^{#3}}}}

\renewcommand{\leq}{\leqslant}

\def\eqdef{\stackrel{\mbox{\tiny def}}{=}}     
\newcommand{\ket}[1]{|\kern.3ex#1\kern.3ex\rangle}
\newcommand{\bra}[1]{\langle\kern.3ex #1 \kern.3ex|}

\newcommand{\re}{\mathop{\mathrm{Re}}\nolimits}      
\newcommand{\im}{\mathop{\mathrm{Im}}\nolimits}      

\def\I{{\rm i}}                  
\def\dd{{\rm d}}                  

\newcommand\antiddots{\mathinner{\mkern2mu\raise1pt\hbox{.}\mkern2mu
    \newline \raise4pt\hbox{.}\mkern2mu\raise7pt\hbox{.}\mkern1mu}}

\newcommand{\abs}[1]{\ensuremath{\left| #1 \right|}}

\def\e{\mathrm{e}}

\newcommand{\moy}[1]{\ensuremath{\left\langle #1 \right\rangle}}

\def\Var{\mathrm{Var}}

\def\O{\mathcal{O}}

\def\T{\mathcal{T}}
\def\Sm{\mathcal{S}}
\def\Qm{\mathcal{Q}}

\def\densG{\rho_{\Gamma}}
\def\densGt{\tilde{\rho}_{\Gamma}}
\def\densT{\rho_t}

\def\D{\mathcal{D}}

\def\rA{r_{\gamma}}

\def\Nt{N_t}

\def\M{{N_{\phi}}}

\usepackage{etoolbox}
\makeatletter
\def\@mkboth#1#2{}
\newlength\appendixwidth
\preto\appendix{\addtocontents{toc}{\protect\patchl@section}}
\newcommand{\patchl@section}{%
  \settowidth{\appendixwidth}{\textbf{Appendix }}%
  \addtolength{\appendixwidth}{1.5em}%
  \patchcmd{\l@section}{1.5em}{\appendixwidth}{}{\ddt}%
}
\makeatother

\begin{document}

\selectlanguage{english}

\title[Distribution of the Wigner-Smith matrix for chaotic cavities
with absorption]{Distribution of the Wigner-Smith time-delay matrix
  for chaotic cavities with absorption and coupled Coulomb gases}

\author{Aur\'elien Grabsch}

\address{Instituut-Lorentz, Universiteit Leiden, P.O. Box 9506, 2300
  RA Leiden, The Netherlands}

\date{\today}

\begin{abstract}
  Within the random matrix theory approach to quantum scattering, we
  derive the distribution of the Wigner-Smith time delay matrix
  $\mathcal{Q}$ for a chaotic cavity with uniform absorption, coupled
  via $N$ perfect channels. In the unitary class $\beta=2$ we obtain a
  compact expression for the distribution of the full matrix in terms
  of a matrix integral. In the other symmetry classes we derive the
  joint distribution of the eigenvalues. We show how the large $N$
  properties of this distribution can be analysed in terms of two
  interacting Coulomb gases living on two different supports. As an
  application of our results, we study the statistical properties of
  the Wigner time delay
  $\tau_{\mathrm{W}} = \mathrm{tr}[\mathcal{Q}]/N$ in the presence of
  absorption.
\end{abstract}

\maketitle

\vspace{10pt}
\noindent\rule{\textwidth}{1pt}
\tableofcontents
\noindent\rule{\textwidth}{1pt}
\vspace{10pt}

\hypersetup{
    linkcolor=Cred
  }


\section{Introduction}
\label{sec:Introduction}

The scattering of waves (quantum or classical) in complex systems has
been a very active field of research, both from the theoretical and
experimental sides. This interest is motivated by applications in
diverse fields, such as nuclear physics\cite{WeiMit09,MitRicWei10},
coherent quantum transport\cite{Bee97}, chaotic
billiards\cite{GuhMulWei98} and propagation of electromagnetic waves
in random media\cite{Sto99}.  In this context, the central object is
the scattering matrix $\Sm(\varepsilon)$ which relates the amplitudes
of incoming and outgoing waves in the different scattering channels at
a given energy $\varepsilon$. The number $N$ of open channels is fixed
by the energy $\varepsilon$ (for example, it is given by transverse
quantisation in a wave guide connected to a cavity). In an ideal
system without losses or gains, the conservation of the particle
number imposes that the scattering matrix is unitary. This matrix can
also satisfy other constraints, depending on the symmetries of the
system. The different symmetry classes are labelled by the Dyson index
$\beta$ \cite{BarMel94,JalPicBee94} (see also the
review\cite{Bee97}). In the absence of time-reversal symmetry
($\beta=2$), the only constraint is the unitarity. If time-reversal
symmetry is preserved ($\beta=1$), $\Sm$ must additionally be
symmetric. The last index $\beta=4$ corresponds to the breaking of
spin-rotation symmetry (in the presence of strong spin-orbit
coupling). In this case, $\Sm$ can be represented by a quaternionic
self-dual unitary matrix.

Another important matrix, which has attracted a lot of attention, is
the Wigner-Smith time delay matrix
$\Qm = -\I \hbar \Sm^\dagger \partial_\varepsilon
\Sm$\cite{Wig55b,Smi60} (in the following we set $\hbar=1$). This
Hermitian matrix contains information about the temporal aspect of the
scattering process. The diagonal elements $\Qm_{ii}$ are called
\textit{injectances} and correspond to the contribution of the
$i^{\mathrm{th}}$ scattering mode to the density of
states\cite{Tex16}. The eigenvalues of $\Qm$, which we denote
$\{ \tau_1, \ldots, \tau_N \}$, are called proper time
delays. Finally, the Wigner time delay, defined as the trace of the
Wigner-Smith matrix,
\begin{equation}
  \label{eq:defWTD}
  \tau_{\rm W} = \frac{1}{N} \tr{\Qm}
  = \frac{1}{N} \sum_{i=1}^N \Qm_{ii}
  = \frac{1}{N} \sum_{i=1}^N \tau_i
  \:,
\end{equation}
plays an important role in many applications, as it is related to the
density of states of the open system (see the review\cite{Tex16}).

For complex systems which exhibit chaotic dynamics, random matrix
theory (RMT) provides a powerful framework to characterise the
statistical properties of the matrices
aforementioned\cite{Bee97,Bro97,Alh00}. The distribution of the
scattering matrix $\Sm(\varepsilon)$ at a given energy $\varepsilon$
has been obtained in the three symmetry classes $\beta=1$, $2$ and $4$
using two different methods: either from a maximal entropy principle
(this is called the stochastic approach\cite{MelPerSel85,MelKum04}),
or by assuming that the Hamiltonian of the closed system can be
described by a random matrix (Hamiltonian
approach\cite{MahWei69,VerWeiZir85}). In the universal regime where
RMT is expected to apply, the two approaches are
equivalent\cite{Bro95}. The resulting distribution, known as the
Poisson kernel, is a cornerstone of the application of RMT to quantum
transport (see the review\cite{Bee97} and references therein).

The Wigner-Smith matrix $\Qm$ is obtained from the energy derivative
of $\Sm$. Therefore it is not sufficient to know the distribution of
$\Sm$ at a given energy: one should also get information about the
energy dependence. Different methods have been introduced to tackle
this more complex
question~\cite{LehSavSokSom95,CopMelBut96,FyoSavSom97,FyoSom97,BroBut97,BroFraBee97,BroFraBee99}.. The
joint distribution of the proper time delays $\{\tau_n\}$, for
perfectly coupled chaotic cavities, has been shown to be related to
the Laguerre ensemble of RMT\cite{BroFraBee97,BroFraBee99},
\begin{equation}
  \label{eq:BFB}
  \mathcal{P} \left( \{ \gamma_i = \tau_{\rm H}/\tau_i \} \right)
  \propto
  \prod_{i<j} \abs{\gamma_i-\gamma_j}^\beta
  \prod_{n=1}^N \gamma_n^{\frac{\beta N}{2}} \:
  \e^{- \frac{\beta}{2} \gamma_n}
  \:,
\end{equation}
where $\tau_{\rm H} = 2\pi / \Delta$ is the Heisenberg time, and
$\Delta$ the mean level spacing of the closed system.  This joint
distribution has been used as a starting point to study many
quantities involving the proper time delays, such as the Wigner time
delay $\tau_{\rm W}$\cite{SavFyoSom01,MezSim13,TexMaj13}.

However in real experiments, absorption is always present to some
level. This leads to losses, which are one source of decoherence in
quantum systems. In particular, the absorption needs to be accounted
for to properly describe the results of some
experiments\cite{DorSmiFre90}. The strength of the absorption is
characterised by the absorption time $\tau_{\rm a}$, which measures
the mean time a wave can spend in the system before being absorbed. It
is convenient to introduce the dimensionless absorption
rate\footnote{In the literature, the dimensionless absorption rate is
  either defined as
  $\tau_{\rm H}/\tau_{\rm a}$\cite{BroBee97,SavSom03} or
  $\tau_{\rm d}/\tau_{\rm a}$\cite{BeeBro01,SavSom04}. The latter
  being more natural to study the limit $N \to \infty$, we prefer it
  here.}  $\gamma = \tau_{\rm d}/\tau_{\rm a}$, where
$\tau_{\rm d} = \tau_{\rm H}/N$ is the dwell time inside the
system. In the following, all the times will be expressed in units of
the Heisenberg time $\tau_{\rm H}$ (i.e. we set $\tau_{\rm H}=1$).

In the presence of absorption, the scattering matrix becomes
sub-unitary. It is thus often referred to as a reflection matrix,
since it encodes the reflection of the fraction of the wave that is
not absorbed by the system. In the following, we will denote this
matrix $\rA$. The Wigner-Smith matrix $\Qm$ in the presence of
absorption measures the deficit of unitarity of the reflection matrix:
$\rA^\dagger \rA = \un_N - \gamma N \Qm$\cite{SavSom03}. These two
matrices are thus related, and one can study either one or the other.

Many results have been obtained on the matrices $\rA$ and $\Qm$ in the
presence of absorption. The joint distribution of the eigenvalues of
$\rA^\dagger \rA$ has been found for $N=1$\cite{BeeBro01} or $N=2$
channels\cite{BroBee97}. For higher number of channels, this
distribution is known only in the limits of strong\cite{KogMelLiq00}
and weak absorption\cite{BeeBro01}. Exact expressions for the mean
density of eigenvalues of $\rA^\dagger \rA$ have been derived for any
number of channels\cite{SavSom03}, and reduce to simpler expressions
in the large $N$ limit\cite{SavSom04}. Concerning the matrix $\rA$
itself, its distribution has been obtained for $N=1$ in the presence
of tunnel coupling\cite{KuhMarMenSto05} or direct
processes\cite{MarMenMar12}. We can also mention that another
important matrix, the Wigner reaction matrix
$K = \I (\rA - \un )/(\rA + \un)$ has been extensively
studied\footnote{In the context of electromagnetic cavities, $K$ is
  related to the impedance matrix of the
  system\cite{ZheAntOtt06,ZheAntOtt06b}.}. The distribution of its
diagonal entries\cite{Fyo03,Fyo04,SavSomFyo05}, and recently the one
of its off-diagonal elements\cite{FedFyo19}, has been found.  For
reviews of the different results and their applications, see for
instance Refs.\cite{FyoSavSom05,FyoSav11}.

Despite all these efforts, the distribution of the matrix $\Qm$ (or
the joint distribution of its eigenvalues) for any absorption rate
$\gamma$ is still unknown. The aim of this paper is to provide this
distribution.

\subsection{Summary of the main results}

Our main results are about the distribution of the Wigner-Smith matrix
$\Qm$ in a chaotic absorbing cavity (with absorption rate $\gamma$),
perfectly coupled to $N$ channels. The distribution is more
conveniently expressed in terms of the inverse matrix
$\Gamma = (N \Qm)^{-1}$ (we rescale by a factor $N$ as the eigenvalues
of $\Qm$ behave as $\O(N^{-1})$ for large $N$).

If time-reversal symmetry is broken (unitary class $\beta=2$), we show
that the distribution of the matrix $\Gamma$ has the compact form
\begin{equation}
  \label{eq:DistrGamBeta2Intro}
  \hspace{-1cm}
  P(\Gamma) \propto
  \e^{- N \tr{\Gamma}}
  \int_0^{\gamma \un_N} \dd T \: 
  \det(\un_{N} \otimes \Gamma -  T \otimes \un_N)
  \:
  \e^{- N \tr{T}}
  \:,
  \qquad
  \Gamma > \gamma \un_N
  \:,
\end{equation}
where the notation indicates that the integral runs over complex
Hermitian matrices $T$ with eigenvalues in $[0, \gamma]$, and
$\otimes$ denotes the Kroenecker product of two matrices. The
eigenvalues of $\Gamma$ are constrained to be larger than the
absorption rate $\gamma$.  This restriction indicates that the
presence of absorption forbids the appearance of large time delays
(small eigenvalues of $\Gamma$) since waves that remain in the system
for too long will be absorbed. In a different context, the
distribution of $\Qm$ for arbitrary tunnel coupling (but no
absorption), was also expressed in terms of an integral over a
$N \times N$ Hermitian matrix\cite{GraSavTex18}.

The distribution~(\ref{eq:DistrGamBeta2Intro}) will be derived in
Section~\ref{sec:UnitCase}, by first obtaining the distribution of the
reflection matrix $\rA$. The method used to obtain this distribution
is difficult to extend to the other symmetry classes due to the
additional constraints satisfied by the matrix $\rA$ when $\beta=1$ or
$4$. We will thus present in Section~\ref{sec:GenCase} a different
derivation, which focuses on the eigenvalues of $\Qm$ and is valid in
the three symmetry classes ($\beta=1$, $2$ or $4$). We obtain the
joint distribution of eigenvalues $\{\Gamma_n\}$ of
$\Gamma = (N\Qm)^{-1}$ as
\begin{eqnarray}
  \label{eq:JPDFGamAbsBetaIntro}
  \fl
  \mathcal{P}(\{ \Gamma_n\}) \propto
  \prod_{i<j} \abs{\Gamma_i-\Gamma_j}^\beta
  \prod_{n=1}^N \e^{- \frac{\beta N}{2} \Gamma_n}
  \\
  \times
  \int_0^\gamma \dd t_1 \cdots \dd t_{\Nt}
  \prod_{i<j}\abs{t_i-t_j}^{\frac{4}{\beta}}
  \prod_{n=1}^{\Nt} \left( [t_n(\gamma-t_n)]^{\frac{2}{\beta}-1}
    \e^{- N t_n}
  \prod_{m=1}^{N}
  \left( \Gamma_m - t_n \right)
    \right)
    \nonumber
  \:,
\end{eqnarray}
where $\Nt = \beta N/2$ and $\Gamma_n > \gamma$. These two results can
be shown to be equivalent for $\beta=2$ by diagonalising the matrices
$\Gamma$ and $T$ in Eq.~(\ref{eq:DistrGamBeta2Intro}). Nevertheless,
we still give the distribution in terms of the full matrix $\Gamma$
for $\beta=2$ as the expression is more compact.


\bigskip

We further show how our
results~(\ref{eq:DistrGamBeta2Intro},\ref{eq:JPDFGamAbsBetaIntro}) can
be used to study the distribution of the Wigner time
delay~(\ref{eq:defWTD}). We develop a modified Coulomb gas technique
to compute the cumulants of $\tau_{\rm W}$ in the limit of large
number $N$ of channels. In the two regimes of weak and strong
absorption, we obtain respectively
\begin{eqnarray}
  \label{eq:WTDweakAbs}
  \moy{\tau_{\rm W}} &\simeq
                       \frac{1}{N}(1 - \gamma)
                       \:,
  &\Var(\tau_W) \simeq
    \frac{4}{\beta N^4} (1-6\gamma)
    \:,
    \quad
    \text{for } \gamma \ll 1
    \:,
  \\
  \label{eq:WTDstrongAbs}
  \moy{\tau_{\rm W}} &\simeq
                       \frac{1}{\gamma N} \left(1 - \frac{1}{\gamma} \right)
                       \:,
                       \quad
  &\Var(\tau_W) \simeq
    \frac{2}{\beta (N \gamma)^4}
    \:,
    \hspace{1.35cm}
    \text{for } \gamma \gg 1
    \:.
\end{eqnarray}
The expansions of the first cumulant in these two limits is consistent
with the known expression $\moy{\tau_{\rm W}} = 1/(N(\gamma+1))$ valid
for any absorption and large number of
channels\cite{SavSom04}. Furthermore, we show that in the regime of
weak absorption, the higher order cumulants can be obtained from the
cumulants at $\gamma=0$ (see Eq.~(\ref{eq:relCumWTD})).


\subsection{Outline of the paper}

The paper is organised as follows. Section~\ref{sec:DistrQ} is mainly
devoted to the derivation of the distribution of the Wigner-Smith
matrix $\Qm$. We first show how to obtain the distribution of the full
matrix in the unitary case ($\beta=2$),
Eq.~(\ref{eq:DistrGamBeta2Intro}), in Section~\ref{sec:UnitCase}. In
Section~\ref{sec:GenCase} we obtain the joint distribution of
eigenvalues~(\ref{eq:JPDFGamAbsBetaIntro}), valid in the three
symmetry classes, starting from the results of
Refs.\cite{VidKan12,JarVidKan15}. We show in Section~\ref{sec:CoulGas}
how the Coulomb gas method can be adapted to handle the
distribution~(\ref{eq:JPDFGamAbsBetaIntro}) in the limit of many open
channels. These results on the distribution of $\Qm$ are used in
Section~\ref{sec:WTD} to study the cumulants of the Wigner time
delay~(\ref{eq:defWTD}) in the presence of absorption.


\section{Distribution of the Wigner-Smith matrix}
\label{sec:DistrQ}


Let us consider a chaotic cavity perfectly coupled to $N$ scattering
channels. In the presence of absorption, with rate $\gamma$, the
$N \times N$ Wigner-Smith matrix $\Qm$ is related to the reflection
matrix $\rA$ as~\cite{SavSom03}
\begin{equation}
  \label{eq:RelReflWS}
  \rA^\dagger \rA = \un_N - \gamma N \Qm
  \:.
\end{equation}
This relation shows that $\Qm$ measures the deficit of unitarity of
the reflection matrix. In particular, when there is no absorption
($\gamma=0$) the reflection matrix becomes unitary.  In the limit of
weak absorption, relation~(\ref{eq:RelReflWS}) has been used to obtain
the distribution of the reflection eigenvalues from the distribution
of $\Qm$ (without absorption)\cite{BeeBro01}. In this paper, we will
follow the opposite route: we will first obtain the distribution of
the reflection matrix for chaotic absorbing cavities, and then deduce
the distribution of $\Qm$ from~(\ref{eq:RelReflWS}).

\begin{figure}
  \centering
  \includegraphics[width=\textwidth]{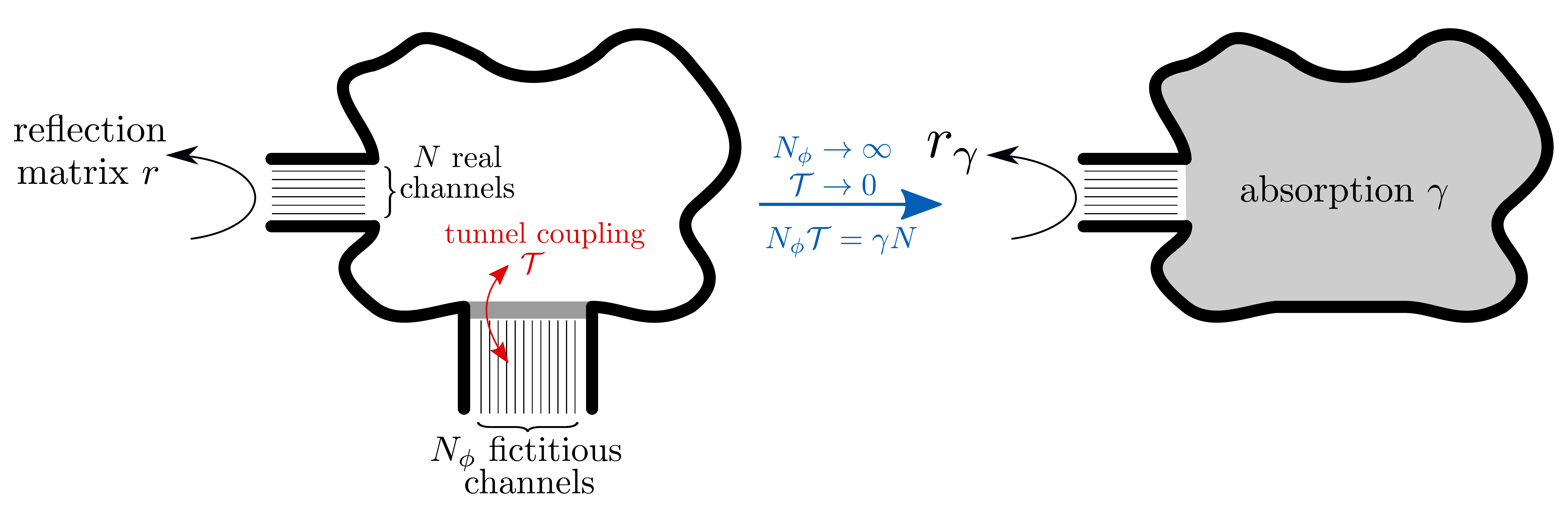}
  \caption{The model for chaotic cavities with absorption. The cavity
    is connected to $N$ real channels via a perfect contact, and to
    $\M$ fictitious channels with tunnel probability $\T$. In the
    limit $\M \to \infty$ and $\T \to 0$ with $\M \T = \gamma N$, this
    model describes a cavity with uniform absorption, with rate
    $\gamma$.}
  \label{fig:CavFicLeads}
\end{figure}

The absorption is modelled by introducing $\M$ fictitious channels,
coupled with tunnel probability $\T$\cite{But86b,BarMel95}. In the
limit of many fictitious channels $\M \to \infty$ and weak coupling $\T
\to 0$ with fixed product
\begin{equation}
  \label{eq:DefTunCoupl}
  \M \T = \gamma N
  \:,
\end{equation}
this model describes a cavity with uniform absorption rate
$\gamma$\footnote{Alternatively one could shift the energy in a model
  \textit{without} absorption along the imaginary axis
  $E_\gamma = E + \I \gamma N/2$ to introduce absorption. The two
  procedures are equivalent\cite{BroBee97}.}\cite{BroBee97}. This is
illustrated in Fig.~\ref{fig:CavFicLeads}. In practice, we will take
this double scaling limit by defining $\T$ in terms of $\M$
using~(\ref{eq:DefTunCoupl}) for $\M \to \infty$ (ensuring that
$\T < 1$). The full system (real and fictitious channels) can be
described by a $(N+\M) \times (N+\M)$ unitary scattering matrix $\Sm$.

Assuming that the dynamics inside the cavity is chaotic, we can follow
the approach of random matrix theory\cite{Bee97,Alh00}. The scattering
matrix $\Sm$ is thus taken as random, with distribution known as the
Poisson kernel\cite{MelPerSel85,Bro95,MelKum04},
\begin{equation}
  \label{eq:PoissonKern}
  P(\Sm) \propto
  \abs{\det(\un - \bar{\Sm}^\star \Sm)}^{-2 - \beta(N+\M-1)}
  \:,
\end{equation}
where $\bar{\Sm}$ is the mean scattering matrix. If we label the first
$N$ lines and columns of $\Sm$ to correspond to the real channels, and
the remaining ones to the $\M$ fictitious channels, $\bar{\Sm}$ takes
the form
\begin{equation}
  \label{eq:Sbar}
  \bar{\Sm} =
  \left(
  \begin{array}{cc}
    0 
    & 0 \\
    0 & \sqrt{1-\T} \: \un_\M
  \end{array}
  \right)
  \:,
\end{equation}
where $\T$ is the tunnel coupling to the fictitious channels. The
zero in the top-left block is a $N \times N$ matrix, which corresponds
to the fact that the real channels are perfectly coupled to the
cavity. We can also decompose the scattering matrix into reflection
and transmission blocks:
\begin{equation}
  \label{eq:DecompSblocks}
  \Sm =
  \left(
  \begin{array}{cc}
    r & t' \\
    t & r'
  \end{array}
  \right)
  \:.
\end{equation}
The $N \times N$ top-left block is the reflection matrix $r$ from the
real channels. In the limit of infinite number of (weakly coupled)
fictitious channels $\M \to \infty$, this block becomes the reflection
matrix of the absorbing cavity:
\begin{equation}
  r \underset{\M \to \infty}{\longrightarrow} \rA
  \:.
\end{equation}
It can then be related to the Wigner-Smith matrix by
Eq.~(\ref{eq:RelReflWS}). Our aim is to obtain the distribution of
this matrix.

In the situation studied in this paper, the $N$ real channels are
equivalent. This means that there is no preferred basis: the matrix
$\Qm$ is invariant under unitary transformations
$\Qm \to U^\dagger \Qm U$ (this is clear on the
distribution~(\ref{eq:DistrGamBeta2Intro}), derived below in
Section~\ref{sec:UnitCase}). The consequence is that the eigenvalues
and eigenvectors of $\Qm$ are statistically uncorrelated, and we can
focus on the eigenvalues $\{ \tau_n \}$ of $\Qm$ only. Equivalently,
thanks to the relation~(\ref{eq:RelReflWS}), we can consider the
eigenvalues $\{ R_n \}$ of $\rA^\dagger \rA$. The joint distribution
of the eigenvalues of $r^\dagger r$ for finite $N$, $\M$ and any tunnel
coupling $\T$ is known\cite{VidKan12,JarVidKan15}. We can thus use
this result to obtain the joint distribution of the reflection
eigenvalues $\{ R_n \}$ in the absorbing situation. This will be done
in Section~\ref{sec:GenCase}. However, we will first present in
Section~\ref{sec:UnitCase} a derivation of the distribution of the
full reflection matrix $r$ in the unitary case $\beta=2$. Besides
providing an alternative derivation to the one given in
Refs.\cite{VidKan12,JarVidKan15}, the procedure described in
Section~\ref{sec:UnitCase} has the advantage to consider the full
reflection matrix $r$ (eigenvalues and eigenvectors), and could in
principle be extended to a situation where the $N$ real channels are
not equivalent.


\subsection{Unitary class}
\label{sec:UnitCase}

Let us first consider the case of broken time-reversal symmetry, which
corresponds to the Dyson index $\beta=2$. We start from the
distribution of the unitary matrix $\Sm$~(\ref{eq:PoissonKern}). In
terms of the block decomposition~(\ref{eq:DecompSblocks}), it becomes
only a function of the $\M \times \M$ bottom-right block $r'$,
\begin{equation}
  \label{eq:DefP0r}
  P(\Sm) \propto
  \abs{\det(\un_\M - \sqrt{1-\T} \: r')}^{-2(N+\M)}
  \eqdef P_0(r')
  \:.
\end{equation}
Our goal is to obtain the distribution of the $N \times N$ reflection
block $r$. It can formally be written as
\begin{equation}
  \label{eq:PrStartBeta2}
  \hspace{-2cm}
  P(r) \propto
  \int \dd r' \dd t \dd t'
  \:
  P_0(r')
  \: \delta(r^\dagger r + t^\dagger t - \un_N)
  \: \delta(r'^\dagger r + t'^\dagger t - \un_\M)
  \: \delta(r^\dagger t' + t^\dagger r')
  \:,
\end{equation}
where the $\delta$-functions impose the unitarity of $\Sm$, and the
integration measures $\dd r'$, $\dd t$ and $\dd t'$ are the Lebesgue
measures over the spaces of $\M \times \M$, $\M \times N$ and $N \times
\M$ complex matrices respectively. For example
\begin{equation}
  \dd r' = \prod_{i=1}^\M \prod_{j=1}^\M
  \dd \re(r'_{ij}) \: \dd \im(r'_{ij})
  \:.
\end{equation}
The idea is to perform all the integrals in~(\ref{eq:PrStartBeta2}) in
order to obtain a form which is convenient to take the limit
$\M \to \infty$.\\

\noindent
\textit{Getting rid of the Dirac delta-functions}\\

The first step to evaluate the integrals in
Eq.~(\ref{eq:PrStartBeta2}) is to perform the integral over the
$N \times \M$ matrix $t'$. Let us make the change of variables
\begin{equation}
  \label{eq:chVartp}
  t' = -(r^\dagger)^{-1} X
  \:,
\end{equation}
where $X$ is the new matrix variable of size $N \times \M$. The
Jacobian of this transformation is\cite{Mathai}
\begin{equation}
  \dd t' = \det(r^\dagger r)^{-\M} \dd X
  \:.
\end{equation}
The integral~(\ref{eq:PrStartBeta2}) thus becomes
\begin{eqnarray}
  \nonumber
  \fl
  P(r) \propto (\det r^\dagger r)^{-\M}
  \int \dd r' \dd t \dd X \:
  P_0(r') \: \delta(r^\dagger r + t^\dagger t - \un_N)
  \\
  \hspace{4.5cm}
  \times \delta(r'^\dagger r' + X^\dagger(r^\dagger r)^{-1}X - \un_\M)
  \: \delta(t^\dagger r'-X)
  \:.
\end{eqnarray}
The last $\delta$-function straightforwardly cancels the integral over
$X$, so we obtain
\begin{equation}
  \fl
  P(r) \propto (\det r^\dagger r)^{-\M}
  \int \dd r' \dd t \:
  P_0(r') \: \delta(r^\dagger r + t^\dagger t - \un_N)
  \: \delta(r'^\dagger( \un_\M
  + t (r^\dagger r)^{-1}t^\dagger) r' - \un_\M)
  \:.
\end{equation}
In the second $\delta$-function appears the matrix
$\un_\M + t (r^\dagger r)^{-1}t^\dagger$, which is Hermitian and
positive (all its eigenvalues are positive). Therefore, we can make
the change of variables
\begin{equation}
  \label{eq:chVarrp}
  r' = (\un_\M + t (r^\dagger r)^{-1}t^\dagger)^{-1/2} Y
  \:.
\end{equation}
The corresponding Jacobian is\cite{Mathai}
\begin{equation}
  \dd r' = \det(\un_\M + t (r^\dagger r)^{-1}t^\dagger)^{-\M} \dd Y
  = \det(\un_N + (r^\dagger r)^{-1} t^\dagger t)^{-\M} \dd Y
  \:,
\end{equation}
where we have used Sylvester's identity. Using also that
\begin{equation}
  (\un_\M + t(r^\dagger r)^{-1} t^\dagger)^{-1}
  = \un_\M - t (r^\dagger r + t^\dagger t)^{-1} t^\dagger
  \:,
\end{equation}
we deduce
\begin{eqnarray}
  \nonumber
  \fl
  P(r) \propto 
  \int \dd Y \dd t \:
  \det(r^\dagger r + t^\dagger t)^{-\M}
  P_0((\un_\M - t(r^\dagger r + t^\dagger t)^{-1} t^\dagger)^{1/2} Y)
  \\
  \hspace{6cm}
  \times
  \delta(r^\dagger r + t^\dagger t - \un_N)
  \: \delta(Y^\dagger Y - \un_\M)
  \:.
\end{eqnarray}
The combination $r^\dagger r + t^\dagger t$ which appear both in the
determinant and in the argument of $P_0$ can be replaced by the
identity thanks to the first $\delta$-function. The second
$\delta$-function imposes that $Y$ is unitary, therefore
\begin{equation}
  \label{eq:Prbeta2Interm}
  P(r) \propto
  \int_{\mathrm{U}(\M)} \dd \mu(Y) \int \dd t
  \:
  P_0((\un_\M - t t^\dagger)^{1/2} Y)
  \: \delta(r^\dagger r + t^\dagger t - \un_N)
  \:,
\end{equation}
where $\dd \mu(Y)$ denotes the Haar measure on the unitary group. This
expression involves both the combination $t^\dagger t$ and
$t t^\dagger$. The first matrix is of size $N \times N$, while the
second has dimension $\M \times \M$. Since we want eventually to take
the limit $\M \to \infty$, we can assume that $\M > N$. The matrix
$t t^\dagger$ thus has the same eigenvalues as $t^\dagger t$, plus a
series of $\M-N$ eigenvalues equal to zero. Therefore, there exists a
unitary matrix $U$ such that
\begin{equation}
  t t^\dagger = U
  \left(
  \begin{array}{cc}
    t^\dagger t & 0 \\
    0 & 0
  \end{array}
  \right)
  U^\dagger
  \:.
\end{equation}
Since $\dd \mu(U^\dagger Y U) = \dd \mu(Y)$, the
integrals~(\ref{eq:Prbeta2Interm}) become
\begin{equation}
  \hspace{-2.3cm}
  P(r) \propto
  \int_{\mathrm{U}(\M)} \dd \mu(Y) \int \dd t
  \:
  P_0 \left(
    \left(
    \begin{array}{cc}
      (\un_N - t^\dagger t)^{1/2} & 0 \\
      0 & \un_{\M-N}
    \end{array}
    \right)
    Y \right)
  \: \delta(r^\dagger r + t^\dagger t - \un_N)
  \:.
\end{equation}
We can now make the last change of variables
\begin{equation}
  \label{eq:ChVartdagt}
  T = t^\dagger t
  \:.
\end{equation}
This change of variables is not one-to-one, as $T$ defines $t$ up to a
unitary matrix. Nevertheless, the Lebesgue measure $\dd t$ can be
expressed in terms of $T$ and a unitary matrix $V$, uniformly
distributed over $\mathrm{U}(\M)$\cite{Mathai} :
\begin{equation}
  \label{eq:JacChVartdagt}
  \dd t = 2^{-N} \det(T)^{\M-N} \dd T \: \dd \mu(V)
  \:.
\end{equation}
Integration over $V$ yields a constant (the volume of the unitary
group), and we can straightforwardly integrate over the Hermitian
matrix $T$ to obtain
\begin{equation}
  \hspace{-1cm}
  P(r) \propto
  \det(\un_N - r^\dagger r)^{\M-N}
  \int_{\mathrm{U}(\M)} \dd \mu(Y)
  \:
  P_0 \left(
    \left(
    \begin{array}{cc}
      (r^\dagger r)^{1/2} & 0 \\
      0 & \un_{\M-N}
    \end{array}
    \right)
    Y \right)
  \:.
\end{equation}
Replacing $P_0$ by its expression~(\ref{eq:DefP0r}) gives
\begin{equation}
  \label{eq:DistrPrUnitInt}
  P(r) \propto
  \det(\un_N - r^\dagger r)^{\M-N}
  \int_{\mathrm{U}(\M)} \frac{\dd \mu(Y)}
  {\abs{\det(\un_\M - A(r) Y)}^{2(N+\M)}}
  \:,
\end{equation}
where we have introduced the Hermitian matrix
\begin{equation}
  \label{eq:defMatA}
  A(r) = \sqrt{1 - \T}
  \left(
  \begin{array}{cc}
    (r^\dagger r)^{1/2} & 0 \\
    0 & \un_{\M-N}
  \end{array}
  \right)
  \:.
\end{equation}
We have reduced the original integral over the $(N+\M)\times(N+\M)$
unitary matrix $\Sm$ to an integral over the $\M\times \M$ unitary
matrix $Y$. However, this expression~(\ref{eq:DistrPrUnitInt}) is not
convenient to take the limit $\M \to \infty$, as the integration domain
depends explicitly on $\M$. We will now evaluate this last integral.\\

\noindent
\textit{Evaluation of the integral over the unitary group}\\

Integrals of the type
\begin{equation}
  \int_{\mathrm{U}(\M)} \frac{\dd \mu(Y)}{\abs{\det(\un_\M - A Y)}^{2n}}
\end{equation}
have been studied in Ref.\cite{FyoKho07}, using the theory of Schur
functions\cite{Mac95}. However, they have been computed for
$0 \leq n \leq \M$, while in Eq.~(\ref{eq:DistrPrUnitInt}) we have
$n = \M+N > \M$. The idea to evaluate this integral in this domain is
given in the Appendix of Ref.\cite{BroBee97}: we make the change of
variables\footnote{This is a well known change of variables in the
  context of quantum scattering, as it is the one that relates the
  scattering matrix of a cavity with perfect couplings (which would be
  here $U$) to the one with arbitrary coupling (here $Y$). The
  couplings are described by the matrix $A$\cite{Bro95}.}
\begin{equation}
  \label{eq:ChgVarUM}
  Y = A - \sqrt{\un-A^2} \: U (1-A U)^{-1} \: \sqrt{\un-A^2}
  \:,
\end{equation}
where $A$ is Hermitian and $U$ is unitary. The Jacobian of the change
of variables~(\ref{eq:ChgVarUM}) is\cite{Mathai,BroBee97}
\begin{equation}
  \dd \mu(Y) \propto
  \det(\un_\M - A^2)^{-\M}
  \abs{\det(\un_\M - A Y)}^{2\M}
  \dd \mu(U)
  \:,
\end{equation}
which cancels out the power $2\M$ in the denominator
of~(\ref{eq:DistrPrUnitInt}). Furthermore, since
\begin{equation}
  \det(\un_\M - A Y) = \det(\un_\M - A^2) \det(\un_\M - A U)^{-1}
  \:,
\end{equation}
the remaining power of the determinant changes sign. Therefore, the
integral in~(\ref{eq:DistrPrUnitInt}) can be expressed as
\begin{eqnarray}
  \nonumber
  \fl
  \int_{\mathrm{U}(\M)} \frac{\dd \mu(Y)}
  {\abs{\det(\un_\M - A Y)}^{2(N+\M)}}
  \propto
  \det(\un_\M - A^2)^{-\M - 2N}
  \\
  \hspace{5cm}
  \times
  \int_{\mathrm{U}(\M)} \dd \mu(U) \:
  \abs{\det(\un_\M - AU)}^{2N}
  \:.
  \label{eq:UnitInt1}
\end{eqnarray}
The integral on the r.h.s has been computed in Ref.\cite{FyoKho07},
with no restriction on the values of $N$ and $\M$:
\begin{equation}
  \label{eq:UnitInt2}
  \hspace{-2.5cm}
  \int_{\mathrm{U}(\M)} \dd \mu(U) \:
  \abs{\det(\un_\M - AU)}^{2N}
  \propto
  \int \dd Z \:
  \det(\un_N + Z^\dagger Z)^{-\M-2N}
  \det(\un + Z^\dagger Z \otimes A^2)
  \:,
\end{equation}
where the integral runs over the $N \times N$ matrix $Z$ with $N^2$
independent complex entries, and $\otimes$ denotes the Kroenecker
product of two matrices. In order to simplify this expression, we
first introduce the matrix $X = Z^\dagger Z$. We can perform this
change of variables in the integral~(\ref{eq:UnitInt2}) similarly as
we did with the matrix $t$\footnote{In~(\ref{eq:JacChVartdagt}), $t$
  was a $\M \times N$ matrix, while $Z$ is now $N \times N$. Therefore
  we must set $\M=N$ in the
  Jacobian~(\ref{eq:JacChVartdagt}).}~(\ref{eq:ChVartdagt},\ref{eq:JacChVartdagt}). We
obtain
\begin{equation}
  \label{eq:UnitInt3}
  \hspace{-2.5cm}
  \int_{\mathrm{U}(\M)} \dd \mu(U) \:
  \abs{\det(\un_\M - AU)}^{2N}
  \propto
  \int \dd X \:
  \det(\un_N + X)^{-\M-2N}
  \det(\un + X \otimes A^2)
  \:,
\end{equation}
where the integral now runs over the Hermitian and positive matrix
$X$.  Combining Eqs.~(\ref{eq:UnitInt1},\ref{eq:UnitInt3}), we can
express the distribution of the reflection
matrix~(\ref{eq:DistrPrUnitInt}) as
\begin{eqnarray}
  \nonumber
  \fl
  P(r) \propto
  \det(\un_N - r^\dagger r)^{\M-N}
  \det(\un_\M - A(r)^2)^{-\M-2N}
  \\
  \hspace{3cm} \times
  \int \dd X
  \det(\un_N + X)^{-\M-2N}
  \det(\un + X \otimes A(r)^2)
  \:.
  \label{eq:DistRfctZ0}
\end{eqnarray}
Substituting the expression of the matrix $A(r)$~(\ref{eq:defMatA}),
this becomes
\begin{eqnarray}
  \label{eq:DistRfctZ}
  \fl
  P(r) \propto \det(\un_N - r^\dagger r)^{-3N}
  \det\left(\un_N +\T \frac{r^\dagger r}{\un_N - r^\dagger r} \right)^{-\M-2N}
  \\
  \hspace{-0.5cm}
  \times
  \int \frac{\dd X}{\det(\un_N + X)^{3N}}
  \det(\un + (1-\T) X \otimes r^\dagger r)
    \det \left(\un -\T \frac{X}{\un_N + X}\right)^{\M-N}
    \:.
    \nonumber
\end{eqnarray}
From this distribution of the reflection matrix $r$, one can deduce
the joint distribution of the eigenvalues of $r^\dagger r$. By also
diagonalising $X = X^\dagger > 0$, one can recover from the
distribution~(\ref{eq:DistRfctZ}) the joint distribution of reflection
eigenvalues derived in\cite{JarVidKan15}. The main difference with our
derivation is that we are dealing with the full reflection matrix, and
not only the eigenvalues of $r^\dagger r$. Besides providing a more
compact expression for the distribution, our approach is also more
natural to analyse the situation where the $N$ channels are not
equivalent (and thus the eigenvalues and eigenvectors no longer
decouple).

The distribution~(\ref{eq:DistRfctZ}) is well suited to derive the
distribution of the reflection matrix $\rA$ in the presence of
absorption. Indeed, the dimension of the integration domain depends
only on the number $N$ of real channels, and not on the number $\M$ of
fictitious channels which we introduced to model the absorption. The
parameter $\M$ only appears in the power of some determinants, and in
the tunnel coupling $\T$~(\ref{eq:DefTunCoupl}). Therefore, we can now
take the limit $\M \to \infty$. Using the famous identity
$\log \det = \mathrm{tr} \: \log$, we deduce
\begin{equation}
  \hspace{-2cm}
  \det\left(\un_N +
    \frac{\gamma N}{\M}
    \frac{r^\dagger r}{\un_N - r^\dagger r} \right)^{-\M-2N}
  \underset{\M \to \infty}{\longrightarrow}
  \e^{- \gamma N \tr \left[\frac{\rA^\dagger \rA}{\un_N - \rA^\dagger \rA}\right]}
  = \e^{\gamma N^2} \e^{-\gamma N \tr \left[(\un_N - \rA^\dagger \rA)^{-1}\right]}
  \:,
\end{equation}
and similarly for the determinant involving $X$. Finally, we obtain
the distribution of the reflection matrix $\rA$ (i.e. the scattering
matrix of the absorbing cavity):
\begin{equation}
  \hspace{-2.5cm}
  P(r) \propto
  \det(\un_N- \rA^\dagger \rA)^{-3N}
  \e^{-\gamma N \tr \left[(\un_N - \rA^\dagger \rA)^{-1}\right]}
  \int \dd X \frac{\det(\un + X \otimes \rA^\dagger \rA)}
  {\det(\un_N+X)^{3N}}
  \e^{- \gamma N \tr \left[\frac{X}{\un_N+X}\right]}
  \:.
\end{equation}
We can further simplify this expression by introducing the variable
\begin{equation}
  T = \frac{X}{\un_N+X}
  \:,
\end{equation}
which is associated to the following Jacobian\cite{Mathai}:
\begin{equation}
  \dd X = \frac{ \dd T }{\det(\un_N-T)^{2N}}
  \:.
\end{equation}
This transformation yields our final result for the reflection matrix:
\begin{eqnarray}
  \nonumber
  \fl
  P(r) \propto
  \det(\un_N- \rA^\dagger \rA)^{-3N}
  \e^{-\gamma N \tr \left[(\un_N - \rA^\dagger \rA)^{-1}\right]}
  \\
  \hspace{3.5cm}
  \times
  \int_0^{\un_N} \dd T
  \:
  \det(\un - T \otimes (\un_N - \rA^\dagger \rA))
  \: \e^{-\gamma N \tr{T}}
    \:,
  \label{eq:DistrReflMatBeta2}
\end{eqnarray}
where the notation indicates that the integration is performed over
Hermitian matrices $T$ with eigenvalues in $[0,1]$. This distribution
is the extension of the uniform distribution of the scattering matrix
for $\gamma=0$ to the absorbing situation $\gamma>0$.

Before using our result~(\ref{eq:DistrReflMatBeta2}) to derive the
distribution of the Wigner-Smith matrix $\Qm$, let us check that this
distribution properly reproduces the different limits which are
known. In the limit of strong absorption $\gamma \to +\infty$, the
exponentials in~(\ref{eq:DistrReflMatBeta2}) strongly suppress the
distribution for $\rA^\dagger \rA$ away from zero. Therefore, we can
drop the contribution of the integral, and expand
$(\un_N - \rA^\dagger \rA)^{-1} \simeq \un_N + \rA^\dagger \rA$. The
distribution of $\rA$ thus reduces to
$P(\rA) \propto \e^{- \gamma N \: \mathrm{tr}(\rA^\dagger \rA)}$,
which coincides with the result of\cite{KogMelLiq00}. In the converse
limit of weak absorption $\gamma \to 0$, the matrix $\rA$ is weakly
sub-unitary, therefore $\rA^\dagger \rA$ is close to the identity. We
can thus drop the integral in the
distribution~(\ref{eq:DistrReflMatBeta2}), and we recover the result
of\cite{BeeBro01}. Finally, for $N=1$ and any absorption rate
$\gamma$, our result reduces to the distribution given in
Ref.\cite{BeeBro01}.\\

\noindent
\textit{Distribution of the Wigner-Smith matrix}\\

The Wigner-Smith matrix $\Qm$ is directly related to the reflection
matrix $\rA$ via~(\ref{eq:RelReflWS}). As in the well studied case
without absorption ($\gamma=0$), the distribution is more conveniently
expressed in terms of the inverse matrix\footnote{We also rescale by a
  factor $N$ for convenience, since the eigenvalues of $\Qm$ scale
  with $N$ as $\O(N^{-1})$.}
\begin{equation}
  \label{eq:defMatGam}
  \Gamma = (N \Qm)^{-1} = \gamma (\un_N - \rA^\dagger \rA)^{-1}
  \:.
\end{equation}
Since $\rA$ is sub-unitary (the absorption causes losses) the
eigenvalues of $\rA^\dagger \rA$ are in $[0,1]$, therefore those of
$\Gamma$ are larger than the absorption rate $\gamma$. This implies
that the eigenvalues of $\Qm$, the proper time delays $\{ \tau_n \}$,
are smaller than $1/(N \gamma)$. The effect of the absorption is thus
to introduce an upper cutoff, which forbids the existence of
arbitrarily large time delays.

From the distribution of the reflection matrix
$\rA$~(\ref{eq:DistrReflMatBeta2}), we can deduce\footnote{For this,
  we need the Jacobian of~(\ref{eq:defMatGam}). First, we introduce
  $R = \rA^\dagger \rA$, which is associated to the Jacobian
  $\dd \rA = 2^{-N} \dd R \dd \mu(U)$, where $U$ is a Haar distributed
  unitary matrix of size $N$ which can be integrated over (see the
  discussion between Eqs.~(\ref{eq:ChVartdagt})
  and~(\ref{eq:JacChVartdagt})). Then we have
  $\Gamma = \gamma (1-R)^{-1}$, thus
  $\dd R \propto (\det \Gamma)^{-2N}\dd \Gamma$\cite{Mathai}.}
\begin{equation}
  \label{eq:DistrGamBeta2}
  \hspace{-1.5cm}
  \boxed{
  P(\Gamma) \propto \e^{- N \tr{\Gamma}}
  \int_0^{\gamma \un_N} \dd T \:
  \det(\un_N \otimes \Gamma - T \otimes \un_N)
  \: \e^{- N \tr{T}}
  \:,
  \qquad
  \Gamma > \gamma \un_N
  \:,
}
\end{equation}
where we have rescaled the matrix $T$ in the integral by a factor
$\gamma$, in order to obtain a more symmetric expression. This
distribution is the first central result of this paper.

First, let us notice that in the limit of no absorption $\gamma \to
0$, the distribution becomes
\begin{equation}
  P(\Gamma) \propto
  \det(\Gamma \otimes \un_N) \: \e^{- N \tr{\Gamma}}
  = (\det \Gamma)^N \: \e^{- N \tr{\Gamma}}
  \:,
\end{equation}
which is the celebrated Wishart-Laguerre distribution of the inverse
Wigner-Smith matrix at zero absorption\cite{BroFraBee97,BroFraBee99}.

The distribution~(\ref{eq:DistrGamBeta2}) is invariant under unitary
transformations $\Gamma \to U \Gamma U^\dagger$, with
$U \in \mathrm{U}(N)$. This is expected since the $N$ real channels
are equivalent, therefore there is no preferred basis. The consequence
is that the eigenvalues and eigenvectors of $\Gamma$ become
statistically independent, and the eigenvectors are uniformly
distributed. We can therefore integrate over the eigenvectors, and
deduce the joint distribution of eigenvalues
$\{ \Gamma_n \}$\footnote{The Jabobian of the eigendecomposition
  involves the Vandermonde determinant
  $\prod_{i<j}\abs{\Gamma_i-\Gamma_j}^\beta$\cite{Mathai}, which is
  well known in RMT\cite{Meh04,For10}.},
\begin{equation}
  \label{eq:jpdfEigBeta2}
  \hspace{-1.5cm}
  \mathcal{P}(\{ \Gamma_n \}) \propto
  \prod_{i<j} (\Gamma_i-\Gamma_j)^2
  \prod_{i=1}^N \e^{- N \Gamma_i}
  \int_0^{\gamma \un_N} \dd T \:
  \prod_{i=1}^N \det(\Gamma_i \un_N - T) \: \e^{-N \tr{T}}
  \:.
\end{equation}
The integral over the matrix $T$ is also invariant under unitary
transformations $T \to U T U^\dagger$, therefore we can also reduce it to
an integral over the eigenvalues $\{t_n\}$ only:
\begin{equation}
  \label{eq:jpdfEigBeta2b}
  \hspace{-2.5cm}
  \mathcal{P}(\{ \Gamma_n \}) \propto
  \prod_{i<j} (\Gamma_i-\Gamma_j)^2
  \prod_{i=1}^N \e^{- N \Gamma_i}
  \int_0^{\gamma} \dd t_1 \cdots \dd t_N \:
  \prod_{i<j} \left( (t_i-t_j)^2
  (\Gamma_i - t_j) \right)
  \prod_{i=1}^N
  \e^{-N \tr{T}}
  \:.
\end{equation}
In this form, the integral over the eigenvalues $\{t_n\}$ can be
performed using Andr\'eief's identity\cite{And83,For18}, which gives
\begin{equation}
  \label{eq:jpdfEigBeta2c}
  \hspace{-2.2cm}
  \mathcal{P}(\{ \Gamma_n \}) \propto
  \prod_{i<j} (\Gamma_i-\Gamma_j)^2
  \prod_{i=1}^N \e^{- N \Gamma_i}
  \:
  \det \left[
    \int_0^{\gamma} \dd t \: t^{n+m-2} \:  \e^{-N t}
    \prod_{i=1}^N (\Gamma_i - t)
  \right]_{1 \leq n,m \leq N}
  \:.
\end{equation}
This expression is useful to obtain exact expressions for the joint
distribution of eigenvalues $\{ \Gamma_n \}$ for small number $N$ of
channels. For example, for $N=1$, this yields
\begin{equation}
  \hspace{-1cm}
  \mathcal{P}(\Gamma) \propto \e^{- \Gamma}
  \int_0^\gamma \dd t \: (\Gamma-t) \: \e^{-t}
  = \e^{-\Gamma}
  \left(
    \Gamma(1-\e^{-\gamma})
    + \e^{-\gamma}(\gamma+1)-1
  \right)
  \:,
\end{equation}
which coincides with the known result\cite{SavSom03}. However,
Eq.~(\ref{eq:jpdfEigBeta2c}) is not well suited to analyse the limit
of large number $N$ of channels. We will present in
Section~\ref{sec:CoulGas} a method, based on the Coulomb gas
technique, which is more convenient in this case. But before that, we
now extend the result~(\ref{eq:jpdfEigBeta2b}) to the other symmetry
classes $\beta=1$ or $4$.


\subsection{General case}
\label{sec:GenCase}

In Section~\ref{sec:UnitCase} we have obtained the distribution of the
Wigner-Smith matrix $\Qm$ in the unitary case. The derivation that we
have presented is difficult to extend to the other symmetry classes
due to the presence of additional constraints on the scattering matrix
(\ref{eq:DecompSblocks}). For instance, if $\beta=1$ the scattering
matrix is symmetric: $\Sm^{\mathrm{T}} = \Sm$. This constraint needs
to be taken into account and leads to additional
complications. Instead, we will follow a different approach, by
focusing on the joint distribution of the eigenvalues of $\Qm$. This
alternative approach is valid in all three symmetry classes. As we
have seen in the case $\beta=2$ discussed above, since the $N$
channels are equivalent, the eigenvectors of $\Qm$ are statistically
independent from the eigenvalues and are uniformly
distributed. Therefore, determining the joint distribution of the
eigenvalues is sufficient to fully characterise the matrix $\Qm$.

Our starting point is the joint distribution of the reflection
eigenvalues $\{ R_n \}_{n=1,\ldots,N}$ for a chaotic cavity perfectly
coupled to $N$ channels and $\M>N$ other channels via a tunnel barrier
$\T$\cite{JarVidKan15}:
\begin{eqnarray}
  \nonumber
  \fl
  \mathcal{P}(\{ R_n\}) \propto
  \prod_{i<j} \abs{R_i-R_j}^\beta
  \prod_{n=1}^N (1-R_n)^{\frac{\beta}{2}-1 + \frac{\beta}{2}(\M-N)}
  (1-(1-\T)R_n)^{-1 - \frac{\beta}{2}(2N+\M-1)}
  \\
  \fl
  \times
  \int_0^1 \dd t_1 \cdots \dd t_{\Nt}
  \prod_{i<j}\abs{t_i-t_j}^{\frac{4}{\beta}}
  \prod_{n=1}^{\Nt} \left( [t_n(1-t_n)]^{\frac{2}{\beta}-1}
  \prod_{m=1}^{\M}
  \left( 1 - t_n (1 - (1-\T) R_m) \right)
  \right)
  \:,
  \label{eq:JPDFrefl0AllBeta}
\end{eqnarray}
where $\Nt = \beta N/2$ and $R_n = 1$ for $n > N$. There is a duality
between the cases $\beta=1$ and $\beta=4$: the distribution of the
reflection eigenvalues for $\beta=1$ is given in terms of an integral
for $\beta=4$, and vice-versa. This type of duality has also been
found for integrals over the Ginibre ensembles\cite{ForRai09}.
Additionally, for $\beta=1$ the dimension $\Nt = N/2$ of the integral
in~(\ref{eq:JPDFrefl0AllBeta}) restricts the number of channels to
even numbers. The problem of finding a similar representation valid
for odd number of channels is still open.

We can obtain the distribution of the reflection eigenvalues in the
presence of absorption from~(\ref{eq:JPDFrefl0AllBeta}) as we did in
the previous section: we set the tunnel coupling $\T = \gamma N/\M$,
Eq.~(\ref{eq:DefTunCoupl}), an let $\M \to \infty$. This gives
\begin{eqnarray}
  \nonumber
  \fl
  \mathcal{P}(\{ R_n\}) \propto
  \prod_{i<j} \abs{R_i-R_j}^\beta
  \prod_{n=1}^N (1-R_n)^{\beta-2 - 3 \beta N/2}
  \:
  \e^{- \frac{\beta N}{2} \gamma \frac{R_n}{1-R_n}}
  \\
  \fl
  \times
  \int_0^1 \dd t_1 \cdots \dd t_{\Nt}
  \prod_{i<j}\abs{t_i-t_j}^{\frac{4}{\beta}}
  \prod_{n=1}^{\Nt} \left( [t_n(1-t_n)]^{\frac{2}{\beta}-1}
    \e^{- \gamma N t_n}
  \prod_{m=1}^{N}
  \left( 1 - t_n (1 -R_m) \right)
  \right)
  \:.
  \label{eq:JPDFreflAbsBeta}
\end{eqnarray}
This is the analogous of Eq.~(\ref{eq:DistrReflMatBeta2}), valid in
the three symmetry classes, but this time expressed in terms of the
eigenvalues of $\rA^\dagger \rA$.

\bigskip

From the joint distribution of the reflection eigenvalues in the
presence of absorption~(\ref{eq:JPDFreflAbsBeta}), we now deduce the
distribution of the eigenvalues of $\Qm$, the proper time delays
$\{ \tau_n \}$. Similarly to the case $\beta=2$ discussed in
Section~\ref{sec:UnitCase}, it is more convenient to work with the
rescaled inverse time delays
\begin{equation}
  \label{eq:RelGamTau}
  \Gamma_n = \frac{1}{N \tau_n} = \frac{\gamma}{1-R_n}
  \:.
\end{equation}
This relation is the analogous of~(\ref{eq:defMatGam}), but expressed
in terms of the eigenvalues. Performing this change of variables in
the distribution~(\ref{eq:JPDFreflAbsBeta}), and rescaling the
integration variables by a factor $\gamma$, we obtain
\begin{eqnarray}
  \nonumber
  \fl
  \mathcal{P}(\{ \Gamma_n\}) \propto
  \prod_{i<j} \abs{\Gamma_i-\Gamma_j}^\beta
  \prod_{n=1}^N \e^{- \frac{\beta N}{2} \Gamma_n}
  \\
  \hspace{-1cm}
  \times
  \int_0^\gamma \dd t_1 \cdots \dd t_{\Nt}
  \prod_{i<j}\abs{t_i-t_j}^{\frac{4}{\beta}}
  \prod_{n=1}^{\Nt} \left( [t_n(\gamma-t_n)]^{\frac{2}{\beta}-1}
    \e^{- N t_n}
    \prod_{m=1}^{N}
    \left( \Gamma_m - t_n \right)
  \right)
  \:.
  \label{eq:JPDFGamAbsBeta}
\end{eqnarray}
This joint distribution is the second central result of this paper.
It is the extension to $\beta=1$ and $\beta=4$ of the
distribution~(\ref{eq:jpdfEigBeta2b}) which we derived above for
$\beta=2$.  In the limit of weak absorption $\gamma \to 0$,
Eq.~(\ref{eq:JPDFGamAbsBeta}) reduces to the well-known
Wishart-Laguerre distribution of the inverse proper time
delays\cite{BroFraBee97,BroFraBee99}, Eq.~(\ref{eq:BFB}), as it
should.


\subsection{Coulomb gas description}
\label{sec:CoulGas}

The representation~(\ref{eq:JPDFGamAbsBeta}) can be used to obtain
exact expressions for the distribution of the time delays in the case
of a few open channels $N = 1, 2 , \ldots$. In the converse situation
of many channels $N \to \infty$, the Coulomb gas method has proved to
be a powerful tool to study different quantities involving the
eigenvalues of random matrices, such as the mean density or linear
statistics (quantities of the form $\sum_n f(\lambda_n)$, where the
$\lambda_n$'s are the eigenvalues of a matrix and $f$ is any given
function, not necessarily linear\footnote{The name linear comes from
  the fact that there are no products of different eigenvalues.}).
We first recall the main ideas of this technique, and show how to
adapt it in the case where the joint distribution of eigenvalues
involves an integral over a domain whose dimension scale as $\O(N)$, as
in~(\ref{eq:JPDFGamAbsBeta}).  This formalism will be useful to study
the statistical properties of the Wigner time-delay~(\ref{eq:defWTD})
(which is a linear statistics) in Section~\ref{sec:WTD}.

The Coulomb gas method has been developed for invariant ensembles of
random matrices, with a joint distribution of eigenvalues
$\{ \lambda_n \}$ of the form
\begin{equation}
  \label{eq:jpdfEigCoul0}
  \mathcal{P}(\lambda_1, \ldots, \lambda_N)
  \propto
  \prod_{i<j} \abs{\lambda_i-\lambda_j}^\beta
  \prod_{i=1}^N \e^{-N V(\lambda_i)}
  \:,
\end{equation}
where $V$ is a function, called the potential, that diverges
sufficiently fast at infinity to ensure that the distribution can be
normalised (the factor $N$ in the exponent ensures that the
eigenvalues are of order $1$ when $N \to \infty$). This function
depends on the ensemble of random matrices under consideration (for
example $V(x) = x^2$ for the Gaussian ensembles). The idea of the
Coulomb gas technique is to write the
distribution~(\ref{eq:jpdfEigCoul0}) as a Gibbs
weight\cite{Dys62I-III}
\begin{equation}
  \label{eq:Gibbs0}
   \mathcal{P}(\lambda_1, \ldots, \lambda_N)
   \propto
   \e^{- \frac{\beta N^2}{2} E_0(\{ \lambda_n \})}
   \:,
\end{equation}
where we introduced
\begin{equation}
  \label{eq:Ener0}
  E_0(\{ \lambda_n \}) = -\frac{1}{N^2}
  \sum_{i\neq j} \ln \abs{\lambda_i - \lambda_j}
  + \frac{1}{N} \sum_{i=1}^N V(\lambda_i)
  \:.
\end{equation}
This function can be interpreted as the energy of a gas of particles
located at positions $\lambda_i$, placed in a potential $V(\lambda)$,
with logarithmic repulsion (hence the name \textit{Coulomb gas}). For
large $N$, the distribution~(\ref{eq:Gibbs0}) becomes peaked near the
minimum of the energy~(\ref{eq:Ener0}). This observation leads to
important simplifications which allow to solve many problems
analytically in this limit. For a recent overview, we refer to the
introduction of Ref.\cite{GraMajTex16} and the PhD
theses\cite{Nad11,Mar15,Gra18}.

\begin{figure}
  \centering
  \includegraphics[width=0.7\textwidth]{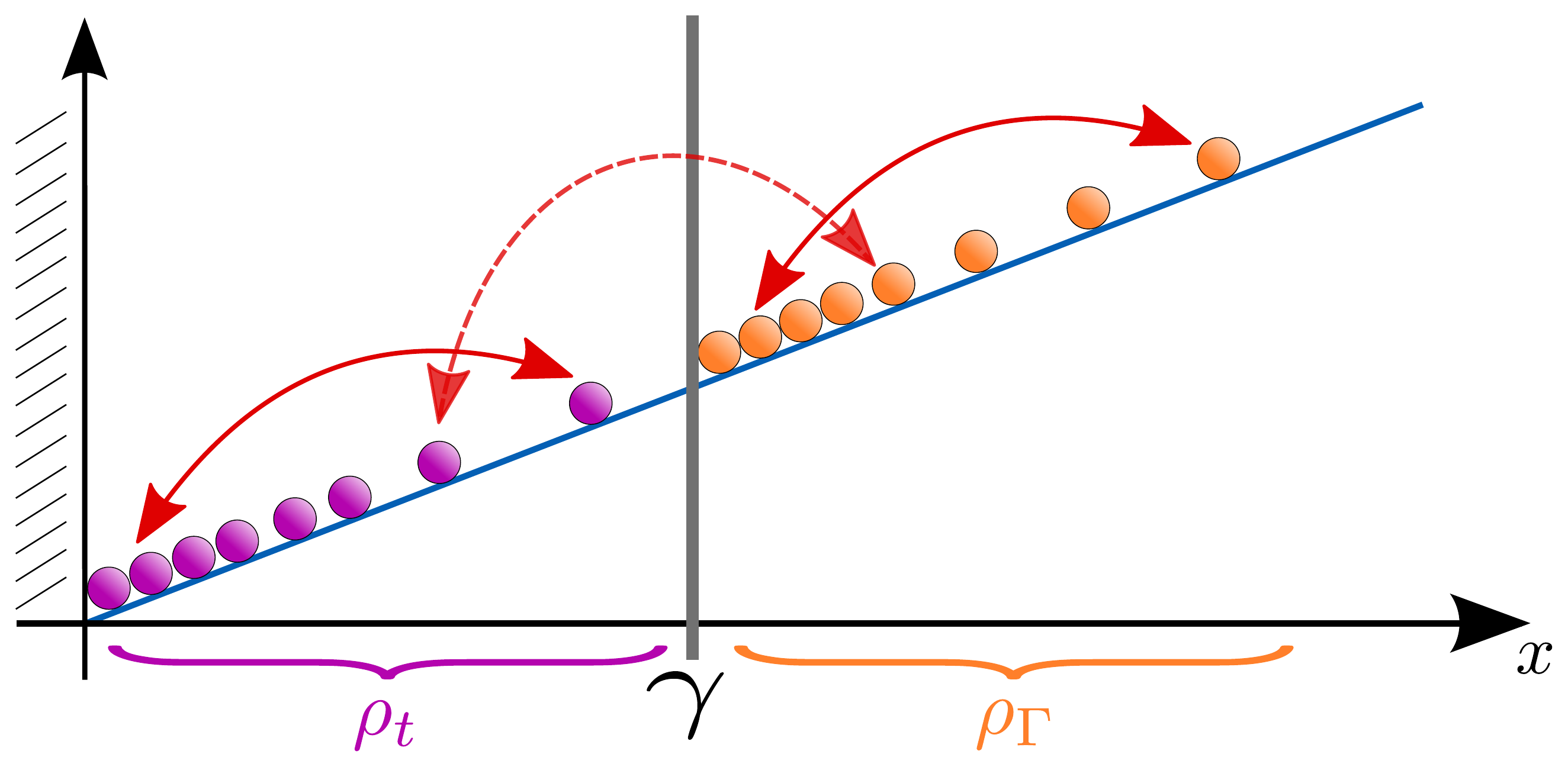}
  \caption{The Coulomb gases associated to the
    energy~(\ref{eq:EnerCoulGas}), or equivalently to the continuous
    version~(\ref{eq:EnerCoulGasCont}). Two log-gases are placed in a
    linear confining potential. The first gas is confined on
    $[0,\gamma]$, while the other one is restricted to
    $[\gamma, +\infty)$. The particles of different gases repel
    logarithmically, with an interaction weaker (dashed arrow) by a
    factor $2$ compared to the repulsion within each gas (solid
    arrows). For large $N$, the two gases can be described by the
    continuous densities $\densT$ and $\densG$.}
  \label{fig:CoulGas}
\end{figure}

In our situation, we cannot apply directly the Coulomb gas method as
described above, since the distribution of eigenvalues
$\Gamma_n = 1/(N \tau_n)$~(\ref{eq:JPDFGamAbsBeta}) involves a
multiple integral over of domain of dimension $\O(N)$. Therefore, we
need to adapt this method by writing the joint
distribution~(\ref{eq:JPDFGamAbsBeta}) in the form
\begin{equation}
  \label{eq:WeightCoul}
  \mathcal{P}(\{ \Gamma_n \}) \propto
  \int_0^\gamma \dd t_1 \cdots \dd t_{\frac{\beta N}{2}}
  \: \e^{- \frac{\beta N^2}{2} E(\{\Gamma_n\}, \{ t_n \})}
  \:,
\end{equation}
where we introduced the energy
\begin{eqnarray}
  \hspace{-2.5cm}
  E(\{\Gamma_n\}, \{ t_n \})
  &=  \frac{1}{N} \sum_{i=1}^N \Gamma_i
    - \frac{1}{N^2} \sum_{i \neq j} \ln \abs{\Gamma_i - \Gamma_j}
    \nonumber
  \\
   &+ \frac{2}{\beta N} \sum_{n=1}^{\beta N/2}
     t_n
     - \frac{4}{\beta^2 N^2} \sum_{n \neq m} \ln \abs{t_i - t_j}
     - \frac{2}{\beta N^2} \left( \frac{2}{\beta} - 1 \right)
     \sum_{n=1}^{\beta N/2} \ln[t_n(\gamma-t_n)]
     \nonumber
  \\
  &- \frac{2}{\beta N^2} \sum_{i=1}^N \sum_{n=1}^{\beta N/2}
    \ln(\Gamma_i - t_n)
    \:.
  \label{eq:EnerCoulGas}
\end{eqnarray}
This expression can be interpreted as the energy of two Coulomb gases:
\begin{itemize}
\item A first gas of $N$ particles, located at positions $\Gamma_i >
  \gamma$. These are the particles of interest, as they are related to
  the eigenvalues of the Wigner-Smith matrix $\Qm$. This gas is placed
  in a linear potential $V(x) = x$ and exhibits logarithmic repulsion
  between the particles (first line of Eq.~(\ref{eq:EnerCoulGas}));
\item A second gas composed of $\beta N/2$ particles, at positions
  $t_i \in [0, \gamma]$. The particles in this gas also repel
  logarithmically, and are placed in the potential
  $\tilde{V}(t) = t - \frac{1}{N}\left(\frac{2}{\beta} - 1\right) \ln
  [t(\gamma-t)]$, which becomes linear for $N \to \infty$ (second line
  of Eq.~(\ref{eq:EnerCoulGas})).
\end{itemize}
These two gases interact with each other logarithmically, but the
strength of this repulsion is half of the one within each gas (see the
discussion in Section~\ref{sec:WTD}). This interpretation is
illustrated in Fig.~\ref{fig:CoulGas}.

\subsubsection*{Continuous formulation}

Instead of working with the sets of eigenvalues, it is more convenient
to introduce the two empirical densities
\begin{equation}
  \label{eq:EmpDens}
  \densG(x) = \frac{1}{N} \sum_{n=1}^N \delta(x - \Gamma_n)
  \quad \text{and} \quad
  \densT(x) = \frac{2}{\beta N} \sum_{n=1}^{\beta N/2}
  \delta(x - t_n)
  \:,
\end{equation}
both normalised to unity. In the limit $N \to \infty$, these densities
can be replaced by continuous ones. Since $t_n \in [0,\gamma]$, the
support of $\densT$ is contained in $[0,\gamma]$. Similarly, the
support of $\densG$ is contained in $[\gamma, +\infty)$. For
$N \gg 1$, the distribution of eigenvalues~(\ref{eq:WeightCoul}) can
be replaced by a weight over the set densities:
\begin{equation}
  \label{eq:DistrGamContCoul}
  \hspace{-2cm}
  \mathcal{P}(\{ \Gamma_n \}) \dd \Gamma_1 \cdots \dd \Gamma_N
  \longrightarrow
  \D \densG
  \:
  \delta \left( \int \densG - 1 \right)
  \int \D \densT \:
  \e^{- \frac{\beta N^2}{2} \mathscr{E}[\densG,\densT]}
  \:
  \delta \left( \int \densT - 1 \right)
  \:,
\end{equation}
where the $\delta$-functions ensure that both densities are
normalised, and the energy functional $\mathscr{E}$ is the continuous
version of~(\ref{eq:EnerCoulGas}):
\begin{eqnarray}
  \mathscr{E}[\densG,\densT]
  &=  \int \dd x \: \densG(x) \: x
    - \int \dd x \dd x' \: \densG(x) \densG(x')
    \ln \abs{x-x'}
    \nonumber
  \\
   &+ \int \dd t \: \densT(t) \: t
    - \int \dd t \dd t' \: \densT(t) \densT(t')
    \ln \abs{t-t'}
     \nonumber
  \\
  &- \int \dd x \dd t \: \densG(x) \densT(t)
    \ln(x - t)
    \:.
  \label{eq:EnerCoulGasCont}
\end{eqnarray}
We have neglected the subleading $\frac{1}{N}$ corrections, and in
particular the entropy which arises when replacing the discrete sets
of eigenvalues by the continuous
densities\cite{Dys62I-III,DeaMaj06,DeaMaj08}. We will now use this
formulation to study the Wigner time delay.


\section{Moments of the Wigner time delay}
\label{sec:WTD}


As an application of our
results~(\ref{eq:DistrGamBeta2},\ref{eq:JPDFGamAbsBeta}) for the
distribution of the Wigner-Smith matrix $\Qm$, we study the
statistical properties of the Wigner time delay
\begin{equation}
  \tau_{\rm W} = \frac{1}{N} \tr{\Qm}
  = \frac{1}{N} \sum_{n=1}^N \tau_n
  = \frac{1}{N^2} \sum_{n=1}^N \frac{1}{\Gamma_n}
\end{equation}
in the presence of absorption. (We have used~(\ref{eq:RelGamTau}) to
express $\tau_{\rm W}$ in terms of the eigenvalues $\{ \Gamma_n \}$.)
Since the eigenvalues $\Gamma_n$ are of order $1$, the Wigner time
delay scales as $N^{-1}$ for large $N$. In order to work with
quantities which do not scale with $N$, we introduce the rescaled
variable
\begin{equation}
  \label{eq:DefRescWTD}
  s = N \tau_{\rm W}
  = \frac{1}{N} \sum_{n=1}^N \frac{1}{\Gamma_n}
  \:.
\end{equation}
We follow an approach similar to Ref.\cite{TexMaj13}, where the
distribution of $\tau_{\rm W}$ was derived for $\gamma=0$ in the
large-$N$ limit. However, instead of the distribution we focus on the
moment generating function of the random
variable~(\ref{eq:DefRescWTD}), at fixed absorption rate $\gamma$,
\begin{equation}
  G_\gamma(\mu) =
  \moy{\e^{- \frac{\beta N^2}{2} \mu s}}
  =
  \int \dd \Gamma_1 \cdots \dd \Gamma_N
  \: \mathcal{P}(\{ \Gamma_n \})
  \: \e^{- \frac{\beta N^2}{2} \frac{\mu}{N} \sum_{n} 1/\Gamma_n}
  \:,
\end{equation}
where $\moy{\cdots}$ denotes the average with respect to the joint
distribution~(\ref{eq:JPDFGamAbsBeta}) and we multiplied the argument
$\mu$ by a factor $\beta N^2/2$ to coincide with the scaling of the
energy. We can replace the integration over the eigenvalues
$\{\Gamma_n\}$ by a functional integral over the density $\densG$, as
prescribed by Eq.~(\ref{eq:DistrGamContCoul}):
\begin{equation}
  \label{eq:DistrWTDInt}
  \hspace{-2cm}
  G_\gamma(\mu) =
  \frac{\displaystyle
    \int 
    \D \densG
    \:
    \delta \left( \int \densG - 1 \right)
    \int \D \densT \:
    \e^{- \frac{\beta N^2}{2}
      \left(
        \mathscr{E}[\densG,\densT]
        + \mu \int \dd x \: \densG(x)/x
      \right)}
    \:
    \delta \left( \int \densT - 1 \right)
  }
  {
    \displaystyle
    \int
    \D \densG
    \:
    \delta \left( \int \densG - 1 \right)
    \int \D \densT \:
    \e^{- \frac{\beta N^2}{2} \mathscr{E}[\densG,\densT]}
    \:
    \delta \left( \int \densT - 1 \right)
  }
  \:,
\end{equation}
where the denominator ensures that $G_\gamma(0)=1$, which follows from
the normalisation of the distribution.  For $N \gg 1$, we can estimate
these integrals by a saddle point method. They are dominated by the
densities $\densG$, $\densT$ which minimise the energy $\mathscr{E}$
under the constraints imposed by the $\delta$-functions. We can find
this minimum by introducing Lagrange multipliers $\mu_0^{(\Gamma)}$
and $\mu_0^{(t)}$. For the numerator, we thus need to find the minimum
of
\begin{equation}
  \fl
  \mathscr{F}[\densG,\densT;\mu] \eqdef
  \mathscr{E}[\densG,\densT]
  + \mu_0^{(t)} \left( \int \densT(t)\dd t - 1  \right)
  + \mu_0^{(\Gamma)} \left( \int \densG(x) \dd x - 1  \right)
  + \mu  \int \frac{\densG(x)}{x} \dd x
  \:.
\end{equation}
This can be done by taking the functional derivatives of this
expression with respect to $\densG(x)$ and $\densT(t)$. This gives two
coupled equations for these densities. As usual in random matrix
theory, it is more convenient to work with the derivatives of these
equations (with respect to $x$ for the equation
${\delta \mathscr{F}}/{\delta \densG(x)} = 0$ and to $t$ for
${\delta \mathscr{F}}/{\delta \densT(t)} = 0$). This gives the set of
two coupled equations:
\begin{eqnarray}
  \label{eq:Eq2CoulGasGam}
  \intVp \dd x' \frac{\densG(x')}{x-x'}
  +\frac{1}{2}
  \int \dd t \frac{\densT(t)}{x-t}
  &= \frac{1}{2} - \frac{\mu}{2 x^2}
  \hspace{1cm} &\text{for } x \in \mathrm{Supp}(\densG) \:,
  \\[0.1cm]
  \label{eq:Eq2CoulGasT}
  \intVp \dd t' \frac{\densT(t')}{t-t'}
  + \frac{1}{2}
  \int \dd x \frac{\densG(x)}{t-x}
  &= \frac{1}{2}
  &\text{for } t \in \mathrm{Supp}(\densT) \:,
\end{eqnarray}
where $\intVp$ denotes a principal value integral. These two equations
can be interpreted as the force balance for the two Coulomb gases. Let
us look for example at Eq.~(\ref{eq:Eq2CoulGasGam}). On the
right-hand-side, the first term is the force coming from the linear
confining potential (this is why it is also present in the second
equation~(\ref{eq:Eq2CoulGasT})). The second term, proportional to the
argument $\mu$ of the generating function, acts as an additional
force. Since we probe the statistics of a
quantity~(\ref{eq:DefRescWTD}) that only involves the eigenvalues
$\Gamma_n$, this term is not present in~(\ref{eq:Eq2CoulGasT}). On the
left-hand-side of~(\ref{eq:Eq2CoulGasGam}), the first term is the
force felt by the particle at point $x$ from the repulsion of all the
other particles in the \textit{same} gas. The second term is the force
felt by the same particle from the repulsion of the \textit{other}
gas. The factor $\frac{1}{2}$ shows explicitly that the inter-gas
interaction is weaker than the intra-gas one. This makes a crucial
difference with the situation previously studied in the literature
where the interaction between the gases is the same as within each
gas, see for
instance\cite{MajNadScaViv09,MajNadScaViv11,MajViv12,GraMajTex16} (in
these papers the interaction is the same since the two gases come from
one global Coulomb gas cut in two parts).

To illustrate the impact of this factor $\frac{1}{2}$, let us look at
the situation $\mu=0$. The solution $\densG$
of~(\ref{eq:Eq2CoulGasGam},\ref{eq:Eq2CoulGasT}) is the typical
density of eigenvalues $\{ \Gamma_n \}$ (which is also the density
that dominates the denominator in~(\ref{eq:DistrWTDInt})). This
density was derived in Ref.\cite{SavSom04}\footnote{The result for any
  finite number of channels $N$ is given in\cite{SavSom03}.}, and is
expressed in terms of a cubic root. On the other hand, equations
similar to~(\ref{eq:Eq2CoulGasGam},\ref{eq:Eq2CoulGasT}), but with the
same factor in front of the two integrals were studied
in\cite{MajViv12} (there the number of eigenvalues of Wishart matrices
larger than $\gamma$ is studied). In this case, the density is
expressed in terms of a square root, as it is often the case in random
matrix theory\cite{Meh04,For10}.

\bigskip

Let us denote $\densG^\star(x;\mu)$ and $\densT^\star(x;\mu)$ the
solutions of the saddle point
equations~(\ref{eq:Eq2CoulGasGam},\ref{eq:Eq2CoulGasT}). We can then
estimate the generating function~(\ref{eq:DistrWTDInt})
as\footnote{The notation
  $X \underset{N\to\infty}{\sim} \e^{- \frac{\beta N^2}{2} Y}$ means
  that $\lim_{N \to \infty} \frac{-2}{\beta N^2} \ln X = Y$. }
\begin{equation}
  \label{eq:LDFgenfct}
  G_\gamma(\mu) \underset{N\to\infty}{\sim}
  \e^{- \frac{\beta N^2}{2} \Phi_\gamma(\mu)}
  \:,
\end{equation}
where
\begin{equation}
  \label{eq:PhiMu}
  \hspace{-1cm}
  \Phi_\gamma(\mu) =
  \mathscr{E}[\densG^\star(x;\mu),\densT^\star(x;\mu)]
  + \mu \int \frac{\dd x}{x} \densG^\star(x;\mu)
  - \mathscr{E}[\densG^\star(x;0),\densT^\star(x;0)]
  \:.
\end{equation}
In this expression, the last term comes from the denominator
of~(\ref{eq:DistrWTDInt}). Equation~(\ref{eq:LDFgenfct}) shows that
the cumulants generating function $G_\gamma(\mu)$ takes a large
deviations form, with a large deviations function $\Phi_\gamma(\mu)$.
In order to compute this function, one should in principle compute the
double integrals in the energy
functional~(\ref{eq:EnerCoulGasCont}). However, it is simpler to use
the thermodynamic identity\cite{GraTex16,Gra18}
\begin{equation}
  \label{eq:ThermoId}
  \dt{\Phi_\gamma}{\mu} = \int \frac{\dd x}{x} \densG^\star(x;\mu)
  \:,
\end{equation}
which is the analogous for the generating function of another
thermodynamic identity introduced in the computation of the
distribution of linear
statistics\cite{GraTex15,CunFacViv16}. This identity allows
us to easily study the cumulants of $s$. Indeed, the cumulant
generating function is
\begin{equation}
  \ln G_\gamma(\mu) = \sum_{k=1}^\infty
  \frac{1}{k!} \left( - \frac{\beta N^2 \mu}{2} \right)^k
  \moy{s^k}_{\rm c}
  \simeq - \frac{\beta N^2}{2} \Phi_\gamma(\mu)
  \:,
\end{equation}
where we have denoted $ \moy{s^k}_{\rm c}$ the $k^{\mathrm{th}}$
cumulant of $s$. The cumulants can thus be obtained by differentiating
$\Phi_\gamma$ with respect to $\mu$. For example, the first two cumulants are
\begin{eqnarray}
  \moy{s}_{\rm c}
  &= \moy{s}
    \simeq
    \left. \dt{\Phi_\gamma}{\mu} \right|_{\mu=0}
    \:,
  \\
  \moy{s^2}_{\rm c}
  &= \Var(s)
    \simeq
    - \frac{2}{\beta N^2}
    \left. \dtn{\Phi_\gamma}{\mu}{2} \right|_{\mu=0}
    \:.
\end{eqnarray}
These derivatives can be conveniently computed from the thermodynamic
identity~(\ref{eq:ThermoId}).

We have reduced the problem of finding the cumulants of the Wigner
time delay to finding the solutions of the saddle point
equations~(\ref{eq:Eq2CoulGasGam},\ref{eq:Eq2CoulGasT}). This is
however a complex problem, in particular due to the difference of
prefactors in front of the integrals (as discussed above). Henceforth,
we will study the two limiting cases of weak and strong absorption, in
which the problem simplifies.

\subsection{Weak absorption}

\begin{figure}
  \centering
  \includegraphics[width=0.45\textwidth]{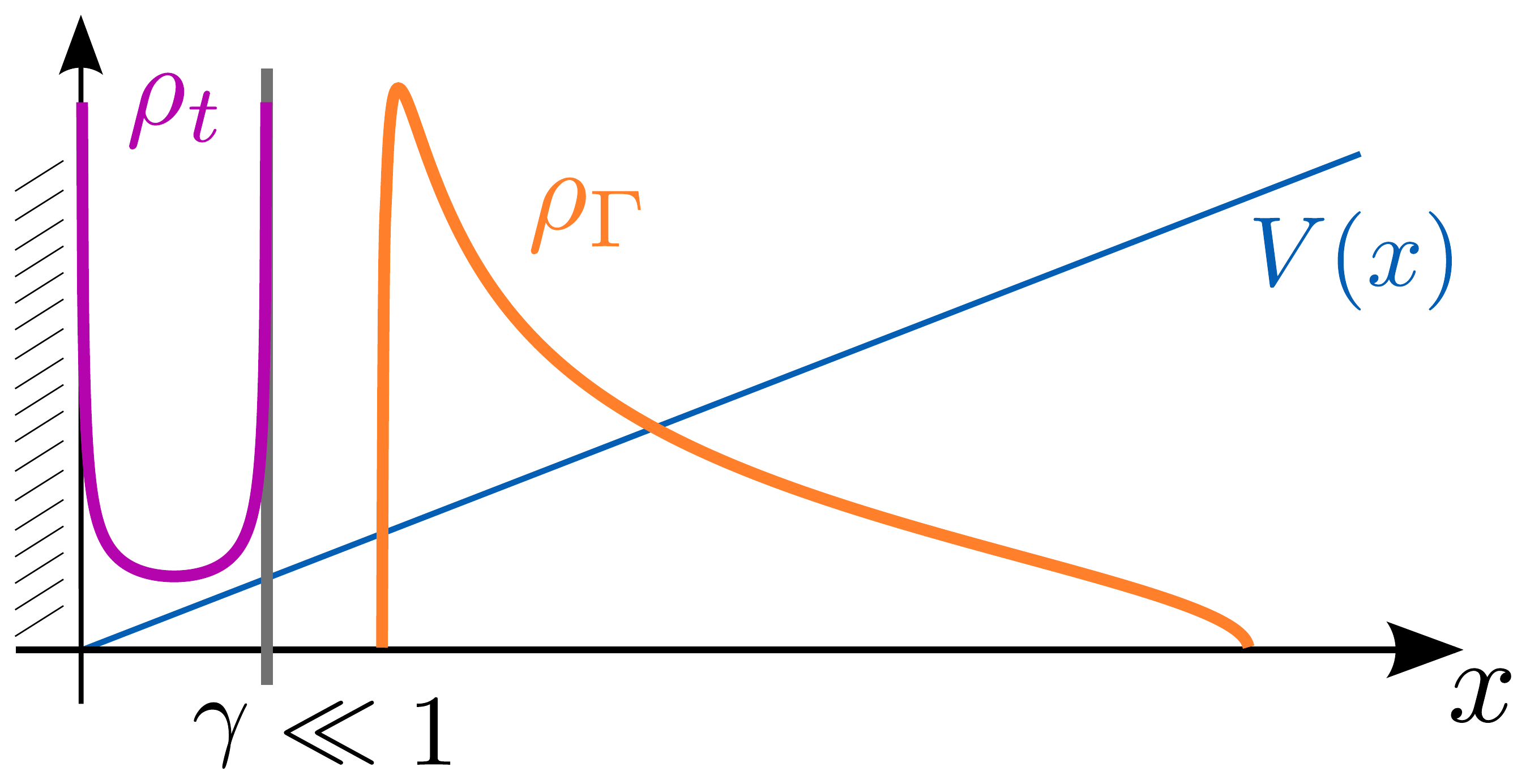}
  \includegraphics[width=0.45\textwidth]{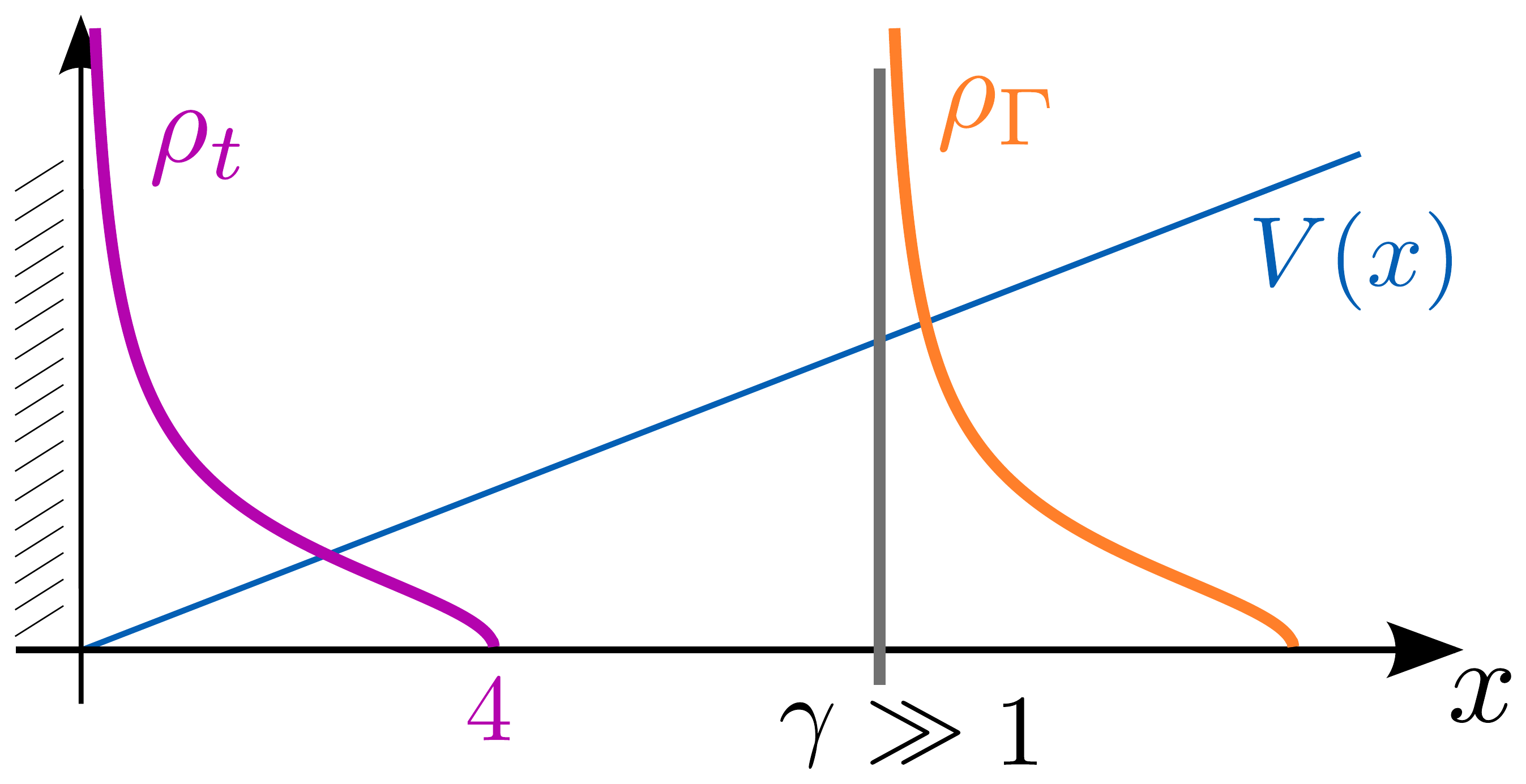}
  \caption{Sketch of the densities of the two Coulomb gases, solutions
    of~(\ref{eq:Eq2CoulGasGam},\ref{eq:Eq2CoulGasT}) in the two
    limiting cases. Left: regime of weak absorption ($\gamma \ll
    1$). Right: regime of strong absorption ($\gamma \gg 1$).}
  \label{fig:CoulGasLimCases}
\end{figure}

Let us first focus on the regime of weak absorption $\gamma \ll 1$.
In this limit the left gas, described by the density $\densT$, is
confined in a small interval $[0,\gamma]$, as illustrated in
Fig.~\ref{fig:CoulGasLimCases} (left). Therefore, the force
balance for this gas~(\ref{eq:Eq2CoulGasT}) is dominated by the
repulsion within the density $\densT$. Both the effect of the other
gas and of the confining potential can thus be neglected. This will
allow us to find the solution $\densT^\star$
of~(\ref{eq:Eq2CoulGasT}), which can then be used to
solve~(\ref{eq:Eq2CoulGasGam}).

Let us formalise this discussion by expanding the force balance
equations~(\ref{eq:Eq2CoulGasGam},\ref{eq:Eq2CoulGasT}) in powers of
$\gamma$. First, we introduce the rescaled density (also normalised to
unity)
\begin{equation}
  \tilde{\densT}(u) =
  \gamma \: \densT(\gamma u)
  \:,
  \qquad
  u \in [0,1]
  \:,
\end{equation}
in terms of which Eq.~(\ref{eq:Eq2CoulGasT}) becomes
\begin{equation}
  \label{eq:Eq2CoulGasTLeadOrder}
  \hspace{-1cm}
  \frac{1}{\gamma}
  \intVp_0^1 \dd u' \frac{\tilde{\densT}(u')}{u-u'}
  = \frac{1}{2} +
  \frac{1}{2} \int \dd x \frac{\densG(x)}{x - \gamma u}
  =
  \frac{\gamma}{2}
  + \frac{\gamma}{2}
  \sum_{n =0}^\infty
  (\gamma u)^n
  \int \dd x \frac{\densG(x)}{x^{n+1}}
  \:.
\end{equation}
We can solve this equation in terms of the constants
$\int \dd x \: \densG(x)/x^{n+1}$, which should be determined
self-consistently later (however as we shall see, these terms will not
contribute at the leading order in $\gamma$). The first of these terms
is $\int \dd x \: \densG(x)/x$, which is the rescaled Wigner time
delay $s$. At leading order in $\gamma$,
Eq.~(\ref{eq:Eq2CoulGasTLeadOrder}) becomes
\begin{equation}
  \intVp_0^1 \dd u' \frac{\tilde{\densT}(u')}{u-u'}
  = \frac{\gamma}{2}(1 + s)
  + \O(\gamma^2)
  \:,
  \qquad
  u \in [0,1]
  \:.
\end{equation}
This integral equation can be solved explicitly by using an inversion
formula due to Tricomi\cite{Tri57}
(see~\ref{sec:TricomiTheorem}). This procedure yields
\begin{equation}
  \tilde{\densT}(u) =
  \frac{1}{\pi \sqrt{u(1-u)}}
  \left[
    1 + \frac{\gamma}{2}(1+s)
    \left(
      \frac{1}{2} - u
    \right)
    + \O(\gamma^2)
  \right]
  \:.
\end{equation}
We can now use this result in the equation for the density
$\densG$~(\ref{eq:Eq2CoulGasGam}), which becomes
\begin{eqnarray}
  \intVp \dd x' \frac{\densG(x')}{x-x'}
  &= \frac{1}{2}
  \left(
    1 - \frac{\mu}{x^2}
    - \int_0^1 \dd u \frac{\tilde{\densT}(u)}{x - \gamma u}
    \right)
    \nonumber
  \\
  &=  \frac{1}{2}
  \left[
    1  - \frac{1}{x}
    - \frac{\mu + \gamma/2}{x^2}
    + \O(\gamma^2)
    \right]
    \:.
\end{eqnarray}
Remarkably, the unknown parameter $s = \int \dd x \: \densG(x)/x$
cancels at first order in $\gamma$. As we could expect, this last
integral equation is similar to the one studied in the case without
absorption ($\gamma=0$)\cite{TexMaj13}. At leading order, the effect
of the absorption is simply to shift the argument $\mu$ of the
generating function by $\gamma/2$. Therefore, we can easily relate the
cumulant generating function for $\gamma>0$ to the one at
$\gamma = 0$,
\begin{equation}
  \Phi_\gamma(\mu) = \Phi_{0}(\mu + \gamma/2)
  + \O(\gamma^2)
  \:,
\end{equation}
which we can equivalently express as
\begin{equation}
  \Phi_\gamma(\mu) = \Phi_{0}(\mu)
  + \frac{\gamma}{2}  \dt{\Phi_0}{\mu} (\mu)
  + \O(\gamma^2)
  \:.
\end{equation}
From this identity between the two generating functions, we can
straightforwardly deduce a relation between the cumulants of the
Wigner time delay with and without absorption:
\begin{equation}
  \label{eq:relCumWTD}
  \boxed{
    \moy{s^n}_{\rm c}
    =  \moy{s^n}_{\rm c}^{(\gamma = 0)}
    - \frac{\gamma \beta N^2}{4}
    \moy{s^{n+1}}_{\rm c}^{(\gamma = 0)}
    + \O(\gamma^2)
    \:.
  }
\end{equation}
We recall that this relation holds for large
$N$. Equation~(\ref{eq:relCumWTD}) expresses the $n^\mathrm{th}$ order
cumulant of the rescaled Wigner time delay at weak absorption
$\gamma > 0$ in terms of the $n^\mathrm{th}$ and $(n+1)^\mathrm{th}$
order cumulants at zero absorption. The cumulants
$\moy{s^n}_{\rm c}^{(\gamma = 0)}$ at zero absorption being
known\cite{MezSim13}, we can straightforwardly apply the
relation~(\ref{eq:relCumWTD}) to deduce the ones in the presence of
weak absorption. For the first two cumulants, this yields the
expressions~(\ref{eq:WTDweakAbs}) given in the introduction.

\subsection{Strong absorption}

We now analyse the regime of strong absorption $\gamma \gg 1$.  In
this case, the right gas is pushed towards the right by the wall at
$x=\gamma$, as shown in Fig.~\ref{fig:CoulGasLimCases} (right). The
interaction between the two gases thus becomes weak compared to the
one within each gas.

To formalise this, let us shift the density $\densG$:
\begin{equation}
  \densGt(x) = \densG(x+\gamma)
  \:.
\end{equation}
Making the substitution $x = \gamma + y$ into the force balance
equation~(\ref{eq:Eq2CoulGasGam}), we obtain
\begin{equation}
  \label{eq:EqDensGStrAbs}
  \intVp \dd y' \frac{\densGt(y')}{y-y'}
  = \frac{1}{2} - \frac{\mu}{2 (y+\gamma)^2}
  - \frac{1}{2}
  \int \dd t \frac{\densT(t)}{y+\gamma-t}
  \:.
\end{equation}
We see on this expression that if we expand in powers of $1/\gamma$,
the term proportional to $\mu$ will be of order $\O(\gamma^{-2})$,
subleading compared to the last term. In order to compensate this
fact, we rescale the parameter $\mu$ by a factor $\gamma$:
\begin{equation}
  \label{eq:RelMuTilde}
  \mu = \gamma \: \tilde{\mu}
  \:.
\end{equation}
In terms of this new parameter, Eq.~(\ref{eq:EqDensGStrAbs}) becomes
\begin{equation}
  \intVp \dd y' \frac{\densGt(y')}{y-y'}
  = \frac{1}{2}
  - \frac{\tilde{\mu}}{2\gamma}
  - \frac{1}{2 \gamma}
  + \O(\gamma^{-2})
  \:,
\end{equation}
where we have used that $\densT$ is normalised to unity.  We can use
Tricomi's theorem (see~\ref{sec:TricomiTheorem}) to solve
this equation. Let us denote $[a,b]$ the support of $\densGt$, we
obtain
\begin{equation}
  \label{eq:DensGtStrAbs}
  \hspace{-1.5cm}
  \densGt(y) = \frac{1}{\pi \sqrt{(y-a)(b-y)}}
  \left[
    1 + \frac{1}{2} \left(
      1 - \frac{\tilde{\mu}+1}{\gamma}
    \right)
    \left(
      \frac{a+b}{2} - y
    \right)
    + \O(\gamma^{-2})
  \right]
  \:.
\end{equation}
Since there is no repulsion from $y=0$, we have $a=0$. The value of
$b$ can be determined by imposing that the density vanishes at
$y=b$. This means that the bracket in~(\ref{eq:DensGtStrAbs}) is zero
for $y=b$. Hence
\begin{equation}
  b = 4 + \frac{4(1+\tilde{\mu})}{\gamma}
  + \O(\gamma^{-2})
  \:,
\end{equation}
and the density~(\ref{eq:DensGtStrAbs}) takes the form
\begin{equation}
  \densGt(y) = \frac{1}{2\pi} \sqrt{\frac{b-y}{y}}
  \left( 1- \frac{\tilde{\mu}+1}{\gamma} + \O(\gamma^{-2}) \right)
  \:.
\end{equation}
From the thermodynamic identity~(\ref{eq:ThermoId}), we deduce the
expression of the cumulant generating function in terms of $\densGt$:
\begin{equation}
  \dt{\Phi_\gamma}{\mu} =
  \frac{1}{\gamma} \dt{\Phi_\gamma}{\tilde{\mu}} =
  \int_0^b \frac{\dd y}{y + \gamma} \densGt(y)
  = \sum_{n=0}^\infty \frac{(-1)^n}{\gamma^{n+1}}
  \int \dd y \: \densGt(y) \: y^n
  \:.
\end{equation}
Since $\densGt$ is normalised to unity, the first term of this series
is $1/\gamma$. We thus have
\begin{equation}
  \frac{1}{\gamma} \dt{\Phi_\gamma}{\tilde{\mu}} =
  \frac{1}{\gamma}
  - \frac{1}{\gamma^2} \int \dd y \: \densGt(y) \: y
  + \frac{1}{\gamma^3} \int \dd y \: \densGt(y) \: y^2
  + \O(\gamma^{-4})
  \:.
\end{equation}
Using the expression of the density $\densGt$~(\ref{eq:DensGtStrAbs}),
we obtain
\begin{equation}
  \frac{1}{\gamma} \dt{\Phi_\gamma}{\tilde{\mu}} =
  \frac{1}{\gamma}
  - \frac{1}{\gamma^2}
  + \frac{1-\tilde{\mu}}{\gamma^3}
  + \O(\gamma^{-4})
  \:.
\end{equation}
From this expansion, we deduce the first two cumulants by taking
derivatives of the generating function,
\begin{equation}
  \moy{s} =
  \left. \dt{\Phi_\gamma}{\mu}
  \right|_{\mu=0}
  =
  \left. \frac{1}{\gamma} \dt{\Phi_\gamma}{\tilde{\mu}}
  \right|_{\tilde{\mu}=0}
  = \frac{1}{\gamma}
  - \frac{1}{\gamma^2}
  + \frac{1}{\gamma^3}
  + \O(\gamma^{-4})
  \:,
\end{equation}
\begin{equation}
  \Var(s) = - \frac{2}{\beta N^2}
  \left. \dtn{\Phi_\gamma}{\mu}{2}
  \right|_{\mu=0}
  =  - \frac{2}{\beta N^2 \gamma^2}
  \left. \dtn{\Phi_\gamma}{\tilde{\mu}}{2}
  \right|_{\tilde{\mu}=0}
  = \frac{2}{\beta N^2 \gamma^4}
  + \O(\gamma^{-5})
  \:,
\end{equation}
which correspond to the expression~(\ref{eq:WTDstrongAbs}) given in
the introduction.


\section{Conclusion}
\label{sec:conclusion}

We have considered the scattering of waves by a chaotic absorbing
cavity, perfectly coupled to $N$ channels. Within the random matrix
theory framework, we have derived the distribution of the Wigner-Smith
time delay matrix $\Qm$ for any absorption rate $\gamma$. This result
thus extends the one of Brouwer, Frahm and
Beenakker\cite{BroFraBee97,BroFraBee99} obtained for zero absorption
$\gamma=0$, to any absorption $\gamma > 0$. Our distribution is
expressed either in terms of an integral over a $N \times N$ positive
Hermitian matrix, or over its eigenvalues. Although providing the
distribution of $\Qm$ in the presence of absorption in most
situations, our derivation should still be extended to yield the
distribution for odd number of channels in the orthogonal class
$\beta=1$.

We have shown how our distribution can be interpreted, in the limit of
many channels $N \to \infty$, in terms of two interacting Coulomb
gases. We have applied this formalism to analyse the cumulants of the
Wigner time delay $\tau_{\rm W} = \tr{\Qm}/N$. In particular, we have
obtained the first cumulants in the two limits of weak
($\gamma \to 0$) and strong ($\gamma \to \infty$)
absorption. Furthermore, we have established a relation between the
cumulants of $\tau_{\rm W}$ at weak absorption and the ones at zero
absorption.

It would be interesting to see if one could derive the expression of
the cumulants of $\tau_{\rm W}$ for any number $N$ of channels from
our new distribution, thus extending the results known at
$\gamma = 0$\cite{MezSim13}. The double Coulomb gas method that we
have introduced in this paper could probably be extended to find the
full distribution of the Wigner time delay in the presence of
absorption (for $N \gg 1$), as it was done for the case without
absorption\cite{TexMaj13}. The situation is however more complex here
as it requires a more detailed analysis of the saddle point
equations~(\ref{eq:Eq2CoulGasGam},\ref{eq:Eq2CoulGasT}). Our technique
could also be used to study other linear statistics involving the
Wigner-Smith matrix, such as trace of higher powers $\tr{\Qm^k}$ and
their correlators, which have been computed for
$\gamma=0$\cite{CunMezSimViv16}.



\section*{Acknowledgements}


\appendix

I am thankful to Yan Fyodorov for pointing out the open question of
the distribution of the Wigner-Smith matrix in the presence of
absorption, and for stimulating discussions. I also thank Christophe
Texier for useful discussions and comments on the manuscript, and
Dmitry Savin for comments on the paper. This project has received
funding from the Netherlands Organization for Scientific Research
(NWO/OCW) and from the European Research Council (ERC) under the
European Union's Horizon 2020 research and innovation programme.


\section{Tricomi's theorem}
\label{sec:TricomiTheorem}

Tricomi's theorem gives an explicit form for the solution of integral
equations of the type
\begin{equation}
  \intVp \dd x' \frac{f(x')}{x-x'}
  = g(x)
  \:,
  \qquad
  x \in \mathrm{Supp}(f)
  \:,
\end{equation}
where $\intVp$ denotes a principal value integral. If we assume that
the support of the solution $f$ has a compact support $[a,b]$, it can
be expressed as\cite{Tri57}
\begin{equation}
  f(x) = \frac{1}{\pi \sqrt{(x-a)(b-x)}}
  \left[
    A + \intVp \frac{\dd t}{\pi}
    \frac{\sqrt{(t-a)(b-t)}}{t-x}
    g(t)
  \right]
  \:,
\end{equation}
where $A = \int_a^b f(x) \dd x$ is a constant. In the situation
considered in this paper, the function $f$ is a density of
eigenvalues, normalised to unity, thus $A=1$.


\section*{References}
\addcontentsline{toc}{section}{References}

\begin{thebibliography}{10}

\bibitem{Alh00}
Y.~Alhassid,
\newblock The statistical theory of quantum dots,
\newblock \href{http://dx.doi.org/10.1103/RevModPhys.72.895}{Rev. Mod.
  Phys.\textbf{ 72}, 895--968} (2000).

\bibitem{And83}
C.~Andr{\'e}ief,
\newblock Note sur une relation entre les int{\'e}grales d{\'e}finies des
  produits des fonctions,
\newblock M{\'e}m. de la Soc. Sci. (Bordeaux)\textbf{ 2}, 1--14 (1886).

\bibitem{BarMel94}
H.~U. Baranger and P.~A. Mello,
\newblock {Mesoscopic transport through chaotic cavities: A random S-matrix
  theory approach},
\newblock \href{http://dx.doi.org/10.1103/PhysRevLett.73.142}{Phys. Rev.
  Lett.\textbf{ 73}, 142--145} (1994).

\bibitem{BarMel95}
H.~U. Baranger and P.~A. Mello,
\newblock Effect of phase breaking on quantum transport through chaotic
  cavities,
\newblock \href{http://dx.doi.org/10.1103/PhysRevB.51.4703}{Phys. Rev.
  B\textbf{ 51}, 4703--4706} (1995).

\bibitem{Bee97}
C.~W.~J. Beenakker,
\newblock Random-matrix theory of quantum transport,
\newblock \href{http://dx.doi.org/10.1103/RevModPhys.69.731}{Rev. Mod.
  Phys.\textbf{ 69}, 731--808} (1997).

\bibitem{BeeBro01}
C.~Beenakker and P.~Brouwer,
\newblock Distribution of the reflection eigenvalues of a weakly absorbing
  chaotic cavity,
\newblock \href{http://dx.doi.org/10.1016/S1386-9477(00)00245-9}{Physica
  E\textbf{ 9}(3), 463 -- 466} (2001).

\bibitem{Bro95}
P.~W. Brouwer,
\newblock Generalized circular ensemble of scattering matrices for a chaotic
  cavity with nonideal leads,
\newblock \href{http://dx.doi.org/10.1103/PhysRevB.51.16878}{Phys. Rev.
  B\textbf{ 51}, 16878--16884} (1995).

\bibitem{BroBee97}
P.~W. Brouwer and C.~W.~J. Beenakker,
\newblock Voltage-probe and imaginary-potential models for dephasing in a
  chaotic quantum dot,
\newblock \href{http://dx.doi.org/10.1103/PhysRevB.55.4695}{Phys. Rev.
  B\textbf{ 55}, 4695--4702} (1997).

\bibitem{BroBut97}
P.~W. Brouwer and M.~B{\"u}ttiker,
\newblock Charge-relaxation and dwell time in the fluctuating admittance of a
  chaotic cavity,
\newblock \href{http://dx.doi.org/10.1209/epl/i1997-00169-0}{Europhys.
  Lett.\textbf{ 37}(7), 441} (1997).

\bibitem{BroFraBee97}
P.~W. Brouwer, K.~M. Frahm, and C.~W.~J. Beenakker,
\newblock Quantum Mechanical Time-Delay Matrix in Chaotic Scattering,
\newblock \href{http://dx.doi.org/10.1103/PhysRevLett.78.4737}{Phys. Rev.
  Lett.\textbf{ 78}, 4737--4740} (1997).

\bibitem{BroFraBee99}
P.~W. Brouwer, K.~M. Frahm, and C.~W.~J. Beenakker,
\newblock {Distribution of the quantum mechanical time-delay matrix for a
  chaotic cavity},
\newblock \href{http://dx.doi.org/10.1088/0959-7174}{Waves in Random
  Media\textbf{ 9}, 91--104} (1999).

\bibitem{Bro97}
P.~Brouwer,
\newblock {\em On the random matrix theory of quantum transport},
\newblock PhD thesis, Leiden University, 1997.

\bibitem{But86b}
M.~B\"uttiker,
\newblock Role of quantum coherence in series resistors,
\newblock \href{http://dx.doi.org/10.1103/PhysRevB.33.3020}{Phys. Rev.
  B\textbf{ 33}, 3020--3026} (1986).

\bibitem{CunFacViv16}
F.~D. Cunden, P.~Facchi, and P.~Vivo,
\newblock {A shortcut through the Coulomb gas method for spectral linear
  statistics on random matrices},
\newblock \href{http://dx.doi.org/10.1088/1751-8113/49/13/135202}{J. Phys. A:
  Math. Theor.\textbf{ 49}(13), 135202} (2016).

\bibitem{CunMezSimViv16}
F.~D. Cunden, F.~Mezzadri, N.~Simm, and P.~Vivo,
\newblock {Correlators for the Wigner-Smith time-delay matrix of chaotic
  cavities},
\newblock \href{http://dx.doi.org/10.1088/1751-8113/49/18/18LT01}{J. Phys. A:
  Math. Theor.\textbf{ 49}(18), 18LT01} (2016).

\bibitem{DeaMaj06}
D.~S. Dean and S.~N. Majumdar,
\newblock {Large Deviations of Extreme Eigenvalues of Random Matrices},
\newblock \href{http://dx.doi.org/10.1103/PhysRevLett.97.160201}{Phys. Rev.
  Lett.\textbf{ 97}, 160201} (2006).

\bibitem{DeaMaj08}
D.~S. Dean and S.~N. Majumdar,
\newblock {Extreme value statistics of eigenvalues of Gaussian random
  matrices},
\newblock \href{http://dx.doi.org/10.1103/PhysRevE.77.041108}{Phys. Rev.
  E\textbf{ 77}, 041108} (2008).

\bibitem{DorSmiFre90}
E.~Doron, U.~Smilansky, and A.~Frenkel,
\newblock Experimental demonstration of chaotic scattering of microwaves,
\newblock \href{http://dx.doi.org/10.1103/PhysRevLett.65.3072}{Phys. Rev.
  Lett.\textbf{ 65}, 3072--3075} (1990).

\bibitem{Dys62I-III}
F.~J. Dyson,
\newblock Statistical Theory of the Energy Levels of Complex Systems. I,
\newblock \href{http://dx.doi.org/10.1063/1.1703773}{J.~Math. Phys.\textbf{
  3}(1), 140--156} (1962),
\newblock {}\\ F. J. Dyson, Statistical Theory of the Energy Levels of Complex
  Systems. II, \href{https://dx.doi.org/10.1063/1.1703774}{J.~Math. Phys.
  \textbf{3}(1), 157--165} (1962),\\ F. J. Dyson,Statistical Theory of the
  Energy Levels of Complex Systems. III,
  \href{https://dx.doi.org/10.1063/1.1703775}{J.~Math. Phys. \textbf{3}(1),
  166--175} (1962).

\bibitem{FedFyo19}
S.~B. Fedeli and Y.~V. Fyodorov,
\newblock {Statistics of off-diagonal entries of Wigner K-matrix for chaotic
  wave systems with absorption},
\newblock \href{https://arxiv.org/abs/1905.04157}{arXiv:1905.04157} (2019).

\bibitem{For10}
P.~J. Forrester,
\newblock {\em Log-gases and random matrices},
\newblock Princeton University Press, 2010.

\bibitem{For18}
P.~J. Forrester,
\newblock {Meet Andréief, Bordeaux 1886, and Andreev, Kharkov 1882–1883},
\newblock \href{http://dx.doi.org/10.1142/S2010326319300018}{Random Matrices
  Theory Appl.\textbf{ 08}(02), 1930001} (2019).

\bibitem{ForRai09}
P.~J. Forrester and E.~M. Rains,
\newblock {Matrix averages relating to Ginibre ensembles},
\newblock \href{http://dx.doi.org/10.1088/1751-8113/42/38/385205}{J. Phys. A:
  Math. Theor.\textbf{ 42}(38), 385205} (2009).

\bibitem{Fyo03}
Y.~V. Fyodorov,
\newblock {Induced vs. Spontaneous breakdown of S-matrix unitarity: Probability
  of no return in quantum chaotic and disordered systems},
\newblock \href{http://dx.doi.org/10.1134/1.1622041}{JETP Letters\textbf{
  78}(4), 250--254} (2003).

\bibitem{FyoSav11}
Y.~V. Fyodorov and D.~V. Savin,
\newblock Resonance scattering of waves in chaotic systems,
\newblock in {\em The Oxford handbook of random matrix theory}, edited by
  G.~Akemann, J.~Baik, and P.~Di~Francesco, pages 703--722, Oxford University
  Press, 2011.

\bibitem{FyoSavSom05}
Y.~V. Fyodorov, D.~V. Savin, and H.-J. Sommers,
\newblock Scattering, reflection and impedance of waves in chaotic and
  disordered systems with absorption,
\newblock \href{http://dx.doi.org/10.1088/0305-4470/38/49/017}{J. Phys.
  A\textbf{ 38}(49), 10731} (2005).

\bibitem{Fyo04}
Y.~V. Fyodorov,
\newblock Complexity of Random Energy Landscapes, Glass Transition, and
  Absolute Value of the Spectral Determinant of Random Matrices,
\newblock \href{http://dx.doi.org/10.1103/PhysRevLett.92.240601}{Phys. Rev.
  Lett.\textbf{ 92}, 240601} (2004).

\bibitem{FyoKho07}
Y.~V. Fyodorov and B.~A. Khoruzhenko,
\newblock {A few remarks on colour{\textendash}flavour transformations,
  truncations of random unitary matrices, Berezin reproducing kernels and
  Selberg-type integrals},
\newblock \href{http://dx.doi.org/10.1088/1751-8113/40/4/007}{J. Phys.
  A\textbf{ 40}(4), 669--699} (2007).

\bibitem{FyoSavSom97}
Y.~V. Fyodorov, D.~V. Savin, and H.-J. Sommers,
\newblock Parametric correlations of phase shifts and statistics of time delays
  in quantum chaotic scattering: Crossover between unitary and orthogonal
  symmetries,
\newblock \href{http://dx.doi.org/10.1103/PhysRevE.55.R4857}{Phys. Rev.
  E\textbf{ 55}, R4857--R4860} (1997).

\bibitem{FyoSom97}
Y.~V. Fyodorov and H.-J. Sommers,
\newblock Statistics of resonance poles, phase shifts and time delays in
  quantum chaotic scattering: Random matrix approach for systems with broken
  time-reversal invariance,
\newblock \href{http://dx.doi.org/10.1063/1.531919}{J. Math. Phys.\textbf{
  38}(4), 1918--1981} (1997).

\bibitem{CopMelBut96}
V.~A. Gopar, P.~A. Mello, and M.~B\"uttiker,
\newblock Mesoscopic Capacitors: A Statistical Analysis,
\newblock \href{http://dx.doi.org/10.1103/PhysRevLett.77.3005}{Phys. Rev.
  Lett.\textbf{ 77}, 3005--3008} (1996).

\bibitem{Gra18}
A.~Grabsch,
\newblock {\em Random matrix theory in statistical physics: quantum scattering
  and disordered systems},
\newblock PhD thesis, Universit\'e Paris Saclay, 2018,
  \href{https://tel.archives-ouvertes.fr/tel-01849097}{https://tel.archives-ouvertes.fr/tel-01849097}.

\bibitem{GraMajTex16}
A.~Grabsch, S.~N. Majumdar, and C.~Texier,
\newblock Truncated Linear Statistics Associated with the Top Eigenvalues of
  Random Matrices,
\newblock \href{http://dx.doi.org/10.1007/s10955-017-1755-5}{J. Stat.
  Phys.\textbf{ 167}(2), 234--259} (2017),
\newblock updated version
  \href{https://arxiv.org/abs/1609.08296}{arXiv:1609.08296}.

\bibitem{GraTex15}
A.~Grabsch and C.~Texier,
\newblock {Capacitance and charge relaxation resistance of chaotic
  cavities---Joint distribution of two linear statistics in the Laguerre
  ensemble of random matrices},
\newblock \href{http://dx.doi.org/10.1209/0295-5075/109/50004}{Europhys.
  Lett.\textbf{ 109}(5), 50004} (2015).

\bibitem{GraTex16}
A.~Grabsch and C.~Texier,
\newblock {Distribution of spectral linear statistics on random matrices beyond
  the large deviation function -- Wigner time delay in multichannel disordered
  wires},
\newblock \href{http://dx.doi.org/10.1088/1751-8113/49/46/465002}{J.~Phys. A:
  Math. Theor.\textbf{ 49}, 465002} (2016).

\bibitem{GraSavTex18}
A.~Grabsch, D.~V. Savin, and C.~Texier,
\newblock {Wigner–Smith time-delay matrix in chaotic cavities with non-ideal
  contacts},
\newblock \href{http://dx.doi.org/10.1088/1751-8121/aada43}{J.~Phys.~A\textbf{
  51}(40), 404001} (2018).

\bibitem{GuhMulWei98}
T.~Guhr, A.~M{\"u}ller-Groeling, and H.~A. Weidenm\"uller,
\newblock Random-matrix theories in quantum physics: common concepts,
\newblock \href{http://dx.doi.org/10.1016/S0370-1573(97)00088-4}{Phys.
  Rep.\textbf{ 299}(4/6), 189--425} (1998).

\bibitem{JalPicBee94}
R.~A. Jalabert, J.-L. Pichard, and C.~W.~J. Beenakker,
\newblock Universal Quantum Signatures of Chaos in Ballistic Transport,
\newblock \href{http://dx.doi.org/10.1209/0295-5075/27/4/001}{Europhys.
  Lett.\textbf{ 27}(4), 255} (1994).

\bibitem{JarVidKan15}
A.~Jarosz, P.~Vidal, and E.~Kanzieper,
\newblock Random matrix theory of quantum transport in chaotic cavities with
  nonideal leads,
\newblock \href{http://dx.doi.org/10.1103/PhysRevB.91.180203}{Phys. Rev.
  B\textbf{ 91}, 180203} (2015).

\bibitem{KogMelLiq00}
E.~Kogan, P.~A. Mello, and H.~Liqun,
\newblock Wave scattering through classically chaotic cavities in the presence
  of absorption: An information-theoretic model,
\newblock \href{http://dx.doi.org/10.1103/PhysRevE.61.R17}{Phys. Rev. E\textbf{
  61}, R17--R20} (2000).

\bibitem{KuhMarMenSto05}
U.~Kuhl, M.~Mart\'{\i}nez-Mares, R.~A. M\'endez-S\'anchez, and H.-J.
  St\"ockmann,
\newblock Direct Processes in Chaotic Microwave Cavities in the Presence of
  Absorption,
\newblock \href{http://dx.doi.org/10.1103/PhysRevLett.94.144101}{Phys. Rev.
  Lett.\textbf{ 94}, 144101} (2005).

\bibitem{LehSavSokSom95}
N.~Lehmann, D.~V. Savin, V.~V. Sokolov, and H.-J. Sommers,
\newblock Time delay correlations in chaotic scattering: random matrix
  approach,
\newblock \href{http://dx.doi.org/10.1016/0167-2789(95)00185-7}{Physica
  D\textbf{ 86}, 572--585} (1995).

\bibitem{Mac95}
I.~G. Macdonald,
\newblock {\em Symmetric functions and Hall polynomials},
\newblock New York: Clarendon, Oxford University Press, 2nd edition, 1995.

\bibitem{MahWei69}
C.~Mahaux and H.~A. Weidenm{\"u}ller,
\newblock {\em Shell-Model Approach to Nuclear Reactions.},
\newblock North-Holland, Amsterdam, 1969.

\bibitem{MajNadScaViv09}
S.~N. Majumdar, C.~Nadal, A.~Scardicchio, and P.~Vivo,
\newblock {Index distribution of Gaussian random matrices},
\newblock \href{http://dx.doi.org/10.1103/PhysRevLett.103.220603}{Phys. Rev.
  Lett.\textbf{ 103}, 220603} (2009).

\bibitem{MajNadScaViv11}
S.~N. Majumdar, C.~Nadal, A.~Scardicchio, and P.~Vivo,
\newblock {How many eigenvalues of a Gaussian random matrix are positive?},
\newblock \href{http://dx.doi.org/10.1103/PhysRevE.83.041105}{Phys. Rev.
  E\textbf{ 83}, 041105} (2011).

\bibitem{MajViv12}
S.~N. Majumdar and P.~Vivo,
\newblock {Number of Relevant Directions in Principal Component Analysis and
  Wishart Random Matrices},
\newblock \href{http://dx.doi.org/10.1103/PhysRevLett.108.200601}{Phys. Rev.
  Lett.\textbf{ 108}, 200601} (2012).

\bibitem{Mar15}
R.~Marino,
\newblock {\em Number statistics in random matrices and applications to quantum
  systems},
\newblock PhD thesis, Universit\'e Paris Saclay, 2015.

\bibitem{MarMenMar12}
A.~M. Mart\'{\i}nez-Arg\"uello, R.~A. M\'endez-S\'anchez, and
  M.~Mart\'{\i}nez-Mares,
\newblock {Wave systems with direct processes and localized losses or gains:
  The nonunitary Poisson kernel},
\newblock \href{http://dx.doi.org/10.1103/PhysRevE.86.016207}{Phys. Rev.
  E\textbf{ 86}, 016207} (2012).

\bibitem{Mathai}
A.~M. Mathai,
\newblock {\em {Jacobians of Matrix Transformations and Functions of Matrix
  Arguments}},
\newblock World Scientific, 1997.

\bibitem{Meh04}
M.~L. Mehta,
\newblock {\em Random matrices},
\newblock Elsevier, Academic, New York, third edition, 2004.

\bibitem{MelKum04}
P.~A. Mello and N.~Kumar,
\newblock {\em Quantum transport in mesoscopic systems -- Complexity and
  statistical fluctuations},
\newblock Oxford University Press, 2004.

\bibitem{MelPerSel85}
P.~A. Mello, P.~Pereyra, and T.~H. Seligman,
\newblock {Information theory and statistical nuclear reactions. I. General
  theory and applications to few-channel problems}SS,
\newblock \href{http://dx.doi.org/10.1016/0003-4916(85)90080-6}{Ann.
  Phys.\textbf{ 161}(2), 254 -- 275} (1985).

\bibitem{MezSim13}
F.~Mezzadri and N.~J. Simm,
\newblock {Tau-Function Theory of Quantum Chaotic Transport with beta=1,2,4},
\newblock \href{http://dx.doi.org/10.1007/s00220-013-1813-z}{Commun. Math.
  Phys.\textbf{ 324}, 465--513} (2013).

\bibitem{MitRicWei10}
G.~E. Mitchell, A.~Richter, and H.~A. Weidenm\"uller,
\newblock Random matrices and chaos in nuclear physics: Nuclear reactions,
\newblock \href{http://dx.doi.org/10.1103/RevModPhys.82.2845}{Rev. Mod.
  Phys.\textbf{ 82}, 2845--2901} (2010).

\bibitem{Nad11}
C.~Nadal,
\newblock {\em {Matrices al{\'e}atoires et leurs applications {\`a} la physique
  statistique et physique quantique}},
\newblock PhD thesis, Universit{\'e} Paris-Sud, 2011.

\bibitem{SavSomFyo05}
D.~V. Savin, H.~J. Sommers, and Y.~V. Fyodorov,
\newblock {Universal statistics of the local Green's function in wave chaotic
  systems with absorption},
\newblock \href{http://dx.doi.org/10.1134/1.2150877}{JETP Letters\textbf{
  82}(8), 544--548} (2005).

\bibitem{SavFyoSom01}
D.~V. Savin, Y.~V. Fyodorov, and H.-J. Sommers,
\newblock Reducing nonideal to ideal coupling in random matrix description of
  chaotic scattering: Application to the time-delay problem,
\newblock \href{http://dx.doi.org/10.1103/PhysRevE.63.035202}{Phys. Rev.
  E\textbf{ 63}, 035202} (2001).

\bibitem{SavSom03}
D.~V. Savin and H.-J. Sommers,
\newblock Delay times and reflection in chaotic cavities with absorption,
\newblock \href{http://dx.doi.org/10.1103/PhysRevE.68.036211}{Phys. Rev.
  E\textbf{ 68}, 036211} (2003).

\bibitem{SavSom04}
D.~V. Savin and H.-J. Sommers,
\newblock Distribution of reflection eigenvalues in many-channel chaotic
  cavities with absorption,
\newblock \href{http://dx.doi.org/10.1103/PhysRevE.69.035201}{Phys. Rev.
  E\textbf{ 69}, 035201} (2004).

\bibitem{Smi60}
F.~T. Smith,
\newblock Lifetime Matrix in Collision Theory,
\newblock \href{http://dx.doi.org/10.1103/PhysRev.118.349}{Phys. Rev.\textbf{
  118}, 349--356} (1960).

\bibitem{Sto99}
H.-J. St{\"o}ckmann,
\newblock {\em Quantum Chaos: An Introduction},
\newblock Cambridge UniversityPress, Cambridge, UK, 1999.

\bibitem{Tex16}
C.~Texier,
\newblock Wigner time delay and related concepts -- Application to transport in
  coherent conductors,
\newblock \href{http://dx.doi.org/10.1016/j.physe.2015.09.041}{Physica
  E\textbf{ 82}, 16--33} (2016),
\newblock Frontiers in quantum electronic transport - In memory of Markus
  Büttiker. See \href{https://arxiv.org/abs/1507.00075}{arXiv:1507.00075} for
  an updated version.

\bibitem{TexMaj13}
C.~Texier and S.~N. Majumdar,
\newblock Wigner Time-Delay Distribution in Chaotic Cavities and Freezing
  Transition,
\newblock \href{http://dx.doi.org/10.1103/PhysRevLett.110.250602}{Phys. Rev.
  Lett.\textbf{ 110}, 250602} (2013).

\bibitem{Tri57}
F.~G. Tricomi,
\newblock {\em {Integral equations}},
\newblock Interscience, London, 1957.

\bibitem{VerWeiZir85}
J.~Verbaarschot, H.~Weidenmüller, and M.~Zirnbauer,
\newblock Grassmann integration in stochastic quantum physics: The case of
  compound-nucleus scattering,
\newblock \href{http://dx.doi.org/10.1016/0370-1573(85)90070-5}{Phys.
  Rep.\textbf{ 129}(6), 367 -- 438} (1985).

\bibitem{VidKan12}
P.~Vidal and E.~Kanzieper,
\newblock Statistics of Reflection Eigenvalues in Chaotic Cavities with
  Nonideal Leads,
\newblock \href{http://dx.doi.org/10.1103/PhysRevLett.108.206806}{Phys. Rev.
  Lett.\textbf{ 108}, 206806} (2012).

\bibitem{WeiMit09}
H.~A. Weidenm\"uller and G.~E. Mitchell,
\newblock Random matrices and chaos in nuclear physics: Nuclear structure,
\newblock \href{http://dx.doi.org/10.1103/RevModPhys.81.539}{Rev. Mod.
  Phys.\textbf{ 81}, 539--589} (2009).

\bibitem{Wig55b}
E.~P. Wigner,
\newblock Lower Limit for the Energy Derivative of the Scattering Phase Shift,
\newblock \href{http://dx.doi.org/10.1103/PhysRev.98.145}{Phys. Rev.\textbf{
  98}, 145--147} (1955).

\bibitem{ZheAntOtt06}
X.~Zheng, T.~M. Antonsen, and E.~Ott,
\newblock Statistics of Impedance and Scattering Matrices in Chaotic Microwave
  Cavities: Single Channel Case,
\newblock
  \href{http://dx.doi.org/10.1080/02726340500214894}{Electromagnetics\textbf{
  26}(1), 3--35} (2006).

\bibitem{ZheAntOtt06b}
X.~Zheng, T.~M. Antonsen, and E.~Ott,
\newblock Statistics of Impedance and Scattering Matrices of Chaotic Microwave
  Cavities with Multiple Ports,
\newblock
  \href{http://dx.doi.org/10.1080/02726340500214902}{Electromagnetics\textbf{
  26}(1), 37--55} (2006).

\end{thebibliography}
\end{document}